\documentclass[reprint,prb,superscriptaddress,amsmath,amssymb,aps
]{revtex4-1}

\usepackage{bm}%
\usepackage[colorlinks=true,linkcolor=blue]{hyperref}%
\expandafter\ifx\csname package@font\endcsname\relax\else
 \expandafter\expandafter
 \expandafter\usepackage
 \expandafter\expandafter
 \expandafter{\csname package@font\endcsname}%
\fi
\usepackage{graphicx}
\usepackage{dcolumn}

\begin{document}

\title{L\'evy flight for electrons in graphene in the presence of regions with enhanced spin-orbit coupling}

\author{Diego B. Fonseca}
\affiliation{Departamento de F\'{\i}sica, Universidade Federal Rural de Pernambuco, Recife - PE, 52171-900, Brazil}
\affiliation{Departamento de F\'{\i}sica, Universidade Federal de Pernambuco, Recife - PE, 50670-901, Brazil}

\author{Anderson L. R. Barbosa}
\email{anderson.barbosa@ufrpe.br}
\affiliation{Departamento de F\'{\i}sica, Universidade Federal Rural de Pernambuco, Recife - PE, 52171-900, Brazil}

\author{Luiz Felipe C. Pereira}
\affiliation{Departamento de F\'{\i}sica, Universidade Federal de Pernambuco, Recife - PE, 50670-901, Brazil}
\affiliation{Dipartimento di Fisica, Sapienza Università di Roma, Roma, 00185, Italy}

\begin{abstract}
In this work, we propose an electronic Lévy glass built from graphene nanoribbons in the presence of regions with enhanced spin-orbit coupling. 
Although electrons in graphene nanoribbons present a low spin-orbit coupling strength, it can be increased by a proximity effect with an appropriate substrate. 
We consider graphene nanoribbons with different edge types, which contain circular regions with a tunable Rashba spin-orbit coupling, whose diameter follow a power-law distribution. 
We find that spin-orbital clusters induce a transition from superdiffusive to diffusive charge transport, similar to what we recently reported for nanoribbons with electrostatic clusters [Phys. Rev. B. 107, 155432 (2023)]. 
We also investigate spin polarization in the spin-orbital Lévy glasses, and show that a finite spin polarization can be found only in the superdiffusive regime.
In contrast, the spin polarization vanishes in the diffusive regime, making the electronic Lévy glass a useful  device whose electronic transmission and spin polarization can be controlled by its Fermi energy.
Finally, we apply a multifractal analysis to charge transmission and spin polarization, and find that the transmission time series in the superdiffusive regime are multifractal, while they tend to be monofractal in the diffusive regime.
In contrast, spin polarization time series are multifractal in both regimes, characterizing a marked difference between mesoscopic fluctuations of charge transport and spin polarization in the proposed electronic L\'evy glass.

\end{abstract}

\maketitle

\section{Introduction} 

With the advent of graphene, two-dimensional materials became one of the most promising research areas in physics and material science \cite{ref1,RevModPhys.92.021003,Sierra_2021,Perkins_2024,D3NH00416C}. This prominence is primarily attributed to their exceptional mechanical, optical, electrical, and magnetic properties.  
Graphene has gained recognition as a promising material for spintronics due to its high electron mobility, low spin-orbit coupling (SOC) strength, negligible hyperfine interaction, and gate tunability \cite{RevModPhys.92.021003,Perkins_2024}.

The key ingredient for spintronic devices and to spintronics in general is the SOC, a relativistic effect in solid-state physics \cite{Bihlmayer2022RashbalikePI,JunsakuNitta2023230102}. 
Spintronics seeks to introduce, control, and identify spins within electronic devices. 
The fundamental requirement for developing a spintronic devices is generating a spin-polarized or purely spin-based current \cite{PhysRevB.99.214446,Liu_2012,Zhang_2014,SANTOS20201,Belayadi_2023,C6CP06972J,10.1063/1.5010973,10.1063/5.0107212,PhysRevB.94.045432,Brede_2023,PhysRevB.93.085408,Park2020ObservationOS}. 
However, in pristine graphene, the intrinsic SOC strength is small, only tens of microelectronvolts \cite{Lu_2023}. 
This can be overcome by depositing graphene on Ni(111) intercalated with a Au layer \cite{PhysRevLett.101.157601}, or by deposition on transition-metal dichalcogenides \cite{RevModPhys.92.021003,Sierra_2021,Perkins_2024}, such as MoS$_2$ or WSe2, and topological insulators \cite{Wang_2015}, such as Bi$_2$Te$_3$, which possess strong SOC. 
The former can increase the extrinsic SOC strength up to 13 meV \cite{PhysRevLett.101.157601}, and in some cases, $\approx$ 100 meV \cite{Marchenko2012GiantRS}. 
The latter give rise to van der Waals heterostructures \cite{RevModPhys.92.021003,Sierra_2021,Perkins_2024}, which, through the proximity-induced SOC, have an active extrinsic SOC strength on the  millielectronvolts scale that can be controlled by a transverse electric field. 

Beyond that, graphene is a unique platform for mimicking wave optics through electronic phenomena. 
This is due to the linear dispersion relation of electrons at low excitation energies, forming the so-called Dirac cone, which aligns qualitatively with the dispersion of photons\cite{Caridad2016}. 
In this context, electronic analogues of some optical phenomena and devices have already been reported. 
For instance, electronic Mie scattering in a graphene ribbon imbibed in a cylindrical electrostatic potential was reported in Refs. [\onlinecite{PhysRevB.87.155409,https://doi.org/10.1002/pssb.201552119}] and experimentally measured in Ref. [\onlinecite{Caridad2016}]. 
Inspired by this experiment, we proposed in Ref. [\onlinecite{PhysRevB.107.155432}] an electronic analogue for an optical Lévy flight device \cite{bart}, dubbed an electronic Lévy glass. 

The optical Lévy glass is built with microspheres whose diameter follow a heavy-tailed distribution, which induces light to have a superdiffusive (or Lévy) dynamic instead of the standard diffusive one\cite{bart}. 
Therefore, the electronic Lévy glass was proposed as a graphene nanoribbon with circular electrostatic clusters whose radii follow a heavy-tailed distribution, which induce electrons to a superdiffusive behavior. 
We showed that the proposed electronic Lévy glass presents a transition from a superdiffusive to a diffusive transport regime, which has not yet been observed in the optical case. 
Since the position of the microspheres and of the electrostatic clusters are fixed in the respective Lévy glasses, subsequent scattering is correlated, in contrast to a typical Lévy walk where the scattering is uncorrelated\cite{bart,PhysRevE.85.021138}.

In general, the transport regime through a Lévy glass can be characterized by the dependence of the average transmission coefficient with the device length $L$, such that \cite{barthelemy2009anomalous,bart,PhysRevE.85.021138}
\begin{equation}\label{eq:alpha}
    \langle T \rangle = \frac{1}{1 + \left(L/\ell\right)^{\gamma}},
\end{equation}
where $\ell$ is the mean free path. 
When $\gamma = 1 $, the usual behavior gives rise to regular diffusive transport. 
Whereas, when $\gamma < 1 $, we have a slow decay of the transmission characterizing a superdiffusive (i.e. Lévy) transport regime. 

\begin{figure}
    \centering
    \includegraphics[clip, width=0.9\linewidth]{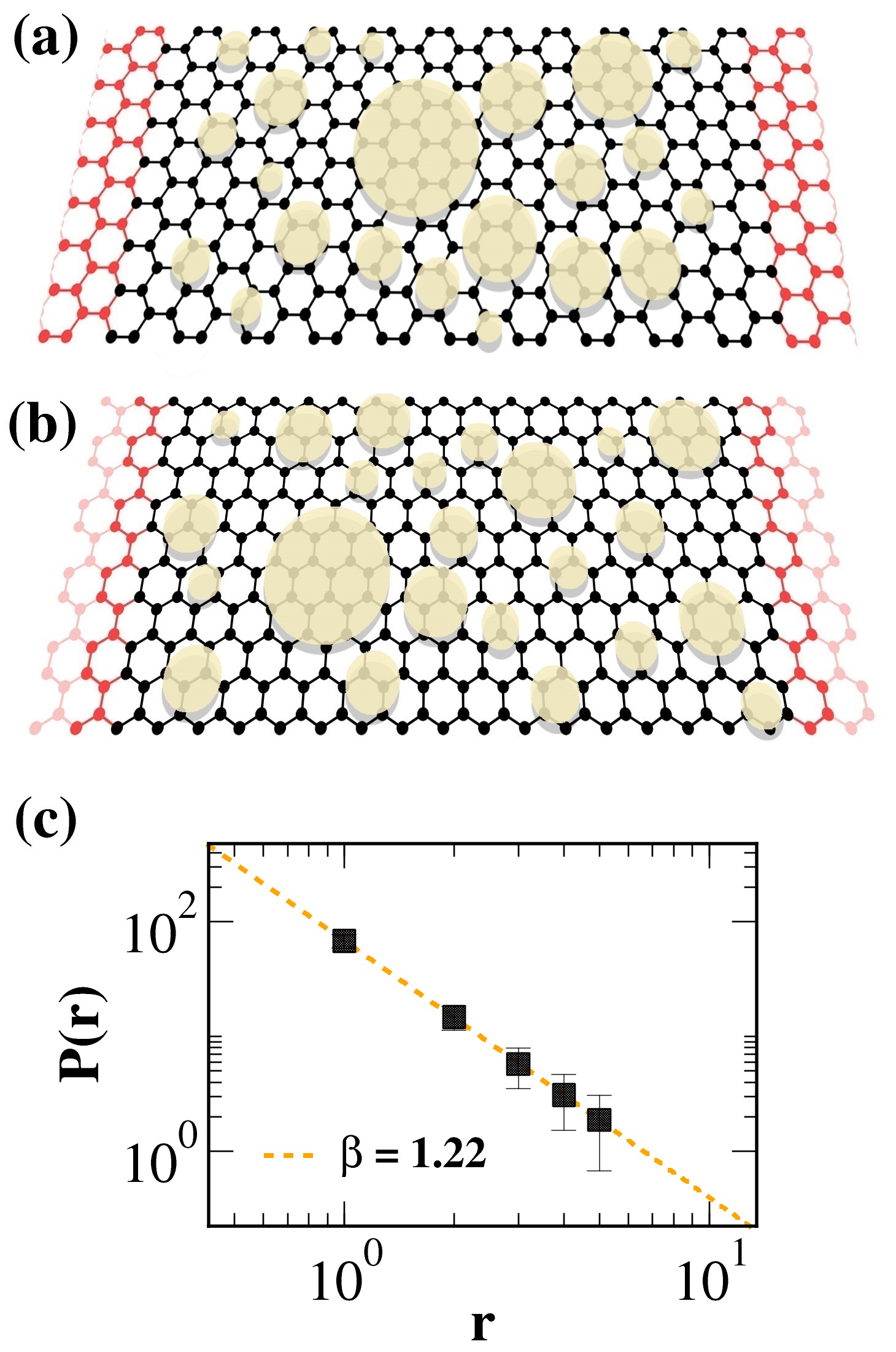}
    \caption{Illustration of (a) AGNR and (b) ZGNR connected to two leads (red). Circular light yellow regions represent the graphene proximity-coupled to a high SOC material. (c) Histogram of cluster radii (symbols); the dashed line is a fit obtained from Eq. (\ref{PR}) with  $\beta = 1.22\pm0.01$.}
    \label{fig:rede_1}
\end{figure}

Intending to extend the application of the  electronic Lévy glass to spintronics, we investigate a spintronic device built from a graphene nanoribbon in the presence of circular clusters with tunable SOC, whose radii follow a power law distribution, as illustrated in Fig. \ref{fig:rede_1}. 
We analyze the impact of the spin-orbital clusters on charge transmission and spin polarization through the Lévy electronic glass using the Landauer-Büttiker formalism. 
Our results show that spin-orbital clusters can induce a transition from superdiffusive to diffusive transport regime as we vary the Fermi energy, similar to the one reported in Ref. [\onlinecite{PhysRevB.107.155432}] with electrostatic clusters. 
Furthermore, we show that an electronic Lévy glass with spin-orbital clusters only presents a finite spin polarization in the superdiffusive regime. 
In contrast, the spin polarization vanishes in the diffusive regime, making this electronic Lévy glass a useful spintronics device.

In order to better understand the superdiffusive-to-diffusive transport transition, we analyze the mesoscopic fluctuations of charge transmission and spin polarization. 
More specifically, we apply a multifractal analysis to charge transmission and spin polarization via a {\it fictional times series}, where the Fermi energy plays the role of {\it a fictional time} \cite{PhysRevLett.79.913,nature,PhysRevE.104.054129,PhysRevLett.128.236803}. 
We find that charge transmission time series in the superdiffusive regime are multifractal, while they tend to be monofractal in the diffusive regime. 
However, between superdiffusive and diffusive regimes, the charge transmission time series shows a significant increase in multifractality, which signals a phase transition \cite{Zhao_2017}. 
Thus, the transition from superdiffusive to diffusive charge transport can be interpreted as a phase transition, probably associated with a chiral symmetry breaking\cite{PhysRevLett.126.206804,PhysRevLett.111.056801}.
In contrast, spin polarization time series are multifractal in both regimes and do not indicate a  phase transition, characterizing a marked difference between mesoscopic fluctuations of charge transport and spin polarization.

This work is organized as follows: in Section \ref{sec:method}, we introduce the graphene tight-binding model, the steps to build the electronic Lévy glass, and a brief review of Multifractal Detrended Fluctuation Analysis \cite{kantelhardt2002multifractal}. 
Section \ref{sec:results} presents our results for charge transmission and spin polarization through the electronic Lévy glass. 
In Section \ref{sec:discussion}, we discuss the origin of the superdiffusive-to-diffusive charge transport transition. 
Finally, we present our conclusions in Section \ref{sec:conslusions}.

\section{Methodology}  \label{sec:method}

We investigate charge transmission and spin polarization through an electronic Lévy glass with spin-orbital clusters. 
The latter is built from a graphene nanoribbon proximity-coupled to a high SOC material, as illustrated in Fig. \ref{fig:rede_1}(a). 
The circular regions illustrate a high SOC material in contact with the graphene nanoribbon. 
Thus, the interface proximity effect induces a Bychkov-Rashba SOC in the circular areas, giving rise to spin-orbital clusters. 
The circle positions are randomly distributed while their radii follow a power-law distribution, as shown in Fig. \ref{fig:rede_1}(c). 

First, we will introduce the microscopic graphene tight-binding model with Bychkov-Rashba SOC, and the steps to build the electronic Lévy glass. 
After that, we briefly present the Multifractal Detrended Fluctuation Analysis (MF-DFA) methodology used to \cite{kantelhardt2002multifractal} analyze the mesoscopic fluctuations of charge transport and spin polarization induced by spin-orbital clusters in the electronic Lévy glass.

\subsection{ Microscopic Model}
Charge transport and spin polarization through graphene nanoribbons in the presence of spin-orbital clusters can be described by the scattering matrix, which is given by  [\onlinecite{datta}]
\begin{equation}
    S = 
    \begin{bmatrix}
       {r} &{t'}\\
        {t} &{r'}
    \end{bmatrix},
\end{equation}
where ${t}({t'})$ and ${r}({r'})$ are the transmission and reflection matrix blocks, respectively. 
The charge transmission coefficient can be calculated from the Landauer-B\"uttiker relation
\begin{equation}
    T = \text{Tr}[{t}{{t}}^\dagger], \label{T}
\end{equation}
which is valid in the linear response regime and for low temperatures. 
In the case of spin-dependent transport, Eq. (\ref{T}) is commonly represented without the trace operator acting on spin space [\onlinecite{Liu_2012}] 
\begin{equation}
T = \begin{bmatrix}
T_{\uparrow\uparrow} & T_{\uparrow\downarrow}\\ 
T_{\downarrow\uparrow} & T_{\downarrow\downarrow},
\end{bmatrix}
\end{equation}
such that it is possible to define individual spin polarizations for each of the directions denoted by the $x,y,$ and $z$-axes
\begin{eqnarray}
    P_{x} &=& \frac{2\text{Re}[T_{\downarrow\uparrow}]}{T_{\uparrow\uparrow} + T_{\downarrow\downarrow}},\nonumber\\
    P_{y} &=& \frac{2\text{Im}[T_{\downarrow\uparrow}]}{T_{\uparrow\uparrow} + T_{\downarrow\downarrow}},\label{Pxyz}\\
    P_{z} &=& \frac{T_{\uparrow\uparrow} - T_{\downarrow\downarrow}}{T_{\uparrow\uparrow} + T_{\downarrow\downarrow}}.\nonumber
\end{eqnarray}

Numerical calculations of the transmission coefficients were performed with KWANT  [\onlinecite{kwant}], which implements a Green’s function–based algorithm within the tight-binding approach. 
The tight-binding Hamiltonian for graphene is given by \cite{PhysRevB.98.045407,PhysRevB.103.L081111,PhysRevB.97.085413}
\begin{equation}
    \hat{H} = -t_0\sum_{\left \langle i,j \right \rangle , \sigma}  c_{i\sigma}^\dagger c_{j\sigma}  
    - \sum_{\langle i,j \rangle, \sigma, \sigma^\prime } \imath \lambda_{i,j} \, c_{i,\sigma}^{\dagger} \left(\left[\textbf{s}\right]_{\sigma \sigma^\prime}\times \hat{\textbf{r}}_{ij}\right)_z c_{j,\sigma^\prime}\,
    \label{H}
\end{equation}
where the indices $i$ and $j$ run over all lattice sites and $\left \langle i,j \right \rangle$ denotes first nearest neighbors, $c_{i,\sigma}$  ($c_{i,\sigma}^\dagger$) are annihilation (creation) operators that remove (add) electrons to site $i$ with spin $\sigma = \uparrow,\downarrow$ and $\hat{\textbf{r}}_{ij}$ is the unit vector along the line segment connecting the sites $i$ and $j$. 
The first term in $\hat{H}$ represents
the usual electron hopping between lattice sites and $t_0$ is the hopping energy, which has a typical value of 2.7 eV [\onlinecite{RevModPhys.81.109}]. 
The second term is the nearest-neighbor hopping term describing the Bychkov-Rashba SOC induced by the interface proximity effect, which explicitly violates $\vec{z} \rightarrow {-}  \vec{z}$ symmetry.
Therefore, the Rashba interaction $\lambda_{ij}$ will be $\lambda_{ij} = \lambda$ when the sites $i$ and $j$ are inside one of the circular regions with spin-orbit coupling, and $\lambda_{ij} = 0$ otherwise.

Figures \ref{fig:rede_1}(a) and \ref{fig:rede_1}(b) depict two electronic Lévy glasses based on armchair (AGNR) and zigzag (ZGNR) graphene nanoribbons in the presence of spin-orbital clusters, which originate from proximity-induced SOC due to a substrate not explicitly shown in the illustration.
The construction of the electronic Lévy glass involves four key steps: (1) random selection of a point on the lattice to be the center of the circular spin-orbital cluster, as well as its radius $R$; (2) assignment of a constant Rashba interaction $\lambda_{ij} = \lambda$ for all neighboring sites $i$ and $j$  within the circular region; (3) random selection of a new lattice point and radius for the next circular region. 
If the circular region overlaps with others already placed, repeat step 3 until a non-overlapping configuration is achieved; then, go to step 2; (4) conclude the procedure after 5,000 consecutive failed attempts to introduce a new circular region. 
We limit the maximum radius of the circular regions to one-eighth of the lattice width without loss of generality. 
Due to the divergence of the second moment of Lévy distributions, introducing a cut-off in the radius of the circular regions is a necessity  [\onlinecite{zaburdaev2015levy}].

Building the graphene nanoribbons with spin-orbital clusters results in a power law radii distribution, as illustrated in Figure \ref{fig:rede_1}(c), which shows the histogram of the radii distribution. 
The probability density $P(r)$ of the random variable $r$ is given by
\begin{equation}
    P(r) \propto \frac{1}{r^{\beta+1}},\label{PR}
\end{equation}
where $0 < \beta < 2$. 
If $0 < \beta < 1$, the first and second moments of $P(r)$ diverge because of heavy tails, while for $1 \le \beta < 2$, only the second moment diverges.
The histogram was obtained from $10^4$ samples of AGNR with width $W_A=49.5a_0$ and length $L_A=51.4a_0$, where $a_0 = 2.49 \text{ \AA}$ is the graphene lattice constant. 
The clusters occupy $42.24\%\pm0.03$ of the lattice area. 
These results remain unchanged for different device lengths $L$ and AGNR or ZGNR widths.

\subsection{Multifractal Analysis} \label{subsec:MFDFA}
In this section, we briefly review the Multifractal Detrended Fluctuation Analysis (MF-DFA), proposed in Ref. [\onlinecite{kantelhardt2002multifractal}] to measure correlations, as well as the multifractal characteristics of charge transmission and spin polarization {\it fictional time series}, where the Fermi energy plays the role of a {\it fictional time}\cite{PhysRevLett.79.913,nature,PhysRevE.104.054129,PhysRevLett.128.236803}. 
Let $T_n \equiv T(n\Delta E) $ be $n$th transmission value (be it charge or spin polarization), where $\Delta E$ is the fictional time step and $n=1,\:2,\:...,\:M$, where $M$  is the number of steps. 
Thus, $E=M\Delta E$ is the range of values for the Fermi energy, and $\{T_n\}$ is the transmission to be analyzed. 
In order to perform the MF-DFA, we first define the profile of $\{T_n\}$ as
\begin{equation}
    \tilde{T}(i)=\sum_{n=1}^i(T_n-\langle T \rangle),
\end{equation}
where $\langle T \rangle$ is the average transmission fictional time series.

We divide the profile series $\{\tilde{T}(i)\}$ into $N_s = \textrm{int}(M/s)$ non-overlapping windows of size $s$, which we denote by $\{\tilde{T}_j(i)\}$, $j=1,\:2,\:...\:N_s$, and fit each window $j$ to a linear function $f_j(i)$.
With the latter, we can compute the variance of each window of the detrended profile series as
\begin{equation}
    F_s^2(j) = \frac{1}{s}\sum_{i=1}^s\{\tilde{T}[(j-1)s+i]-f_j(i)\}^2, \label{Fs}
\end{equation}
where $j=1,\:2,\:...,\:N_s$, and carry out a detrending in the window by fitting the corresponding interval to a linear function $f_j(i)$, $j=1,\:2,\:...,\:N_s$.
Finally, we compute the $q$-th order fluctuation function as
\begin{equation}
    F_q(s)= \left(\frac{1}{2N_s}\sum_{j=1}^{2N_s}\left[F^2_s(j)\right]^{q/2}\right)^{1/q} \label{Fp}
\end{equation}
for a set of real values of $q$.
Once we have obtained the set of functions $F_q(s)$, we study their scaling behavior with the window size $s$ according to the relation
\begin{equation}
    F_q(s) \sim s^{h(q)},\label{hp}
\end{equation}
where $h(q)$ is the generalized Hurst exponent. 
If $h(q)$ is $q$-dependent, then the corresponding time series is said to be multifractal, while if $h(q)$ is independent of $q$, the time series is  monofractal.
We also define
\begin{equation}
    \tau(q)=qh(q)-1,
\end{equation}
such that the multifractal singularity spectrum $f (\alpha)$ is given by a Legendre transform of $\tau(q)$, defined as 
\begin{equation}
    f(\alpha)=\alpha q -\tau(q),\label{fa}
\end{equation}
where $\alpha=d\tau/ d q$. 
From the point of view of the singularity spectrum $f(\alpha)$, we know that multifractal time series are characterized by a broad $f(\alpha)$, while monofractal ones present narrow $f(\alpha)$. 
In other words, the strength of the multifractality can be inferred from the width of $f(\alpha)$, $\Delta\alpha=\alpha_{max} - \alpha_{min}$, such that as $\Delta\alpha\rightarrow 0$, we have a loss of multifractality and a consequent tendency to monofractal behavior.

\section{Results}\label{sec:results}

\begin{figure}
    \centering
    \includegraphics[width=1.0\linewidth]{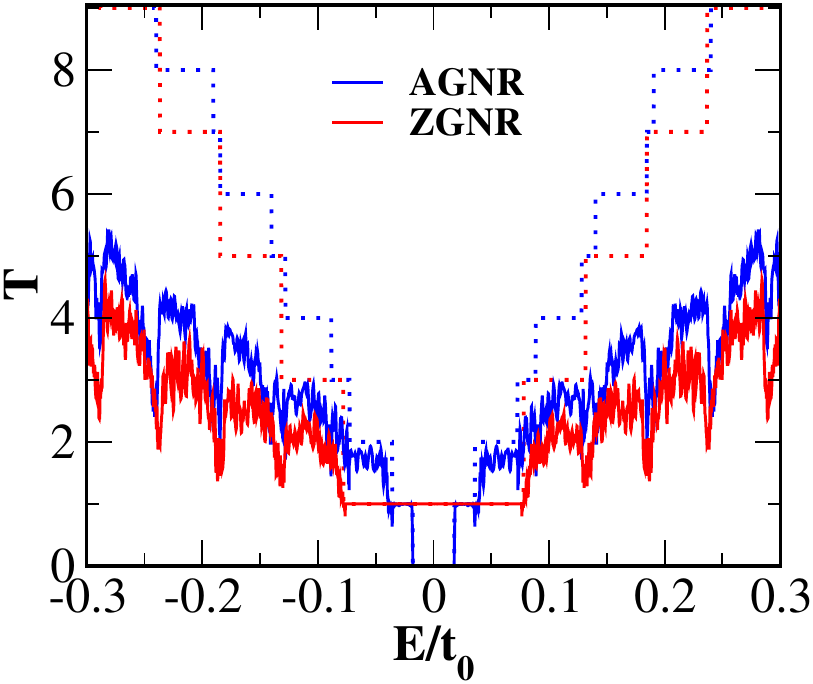}
    \caption{Electronic charge transmission for pristine AGNR and ZGNR (dotted lines), and
     through electronic L\'evy glass with Rashba interaction $\lambda = 0.07t_0$ (continuous lines), as a function of Fermi energy.}
    \label{fig:T_E}
\end{figure}

Figure \ref{fig:T_E} shows the electronic charge transmission,  calculated from Eq. (\ref{T}), as a function of Fermi energy. 
Dotted lines represent the transmission through pristine AGNR and ZGNR of width $W_A = 49.5a_0$ and $W_Z = 49.6a_0$ and length $L_A = 1050.7a_0$ and $L_Z = 1050.5a_0$, respectively. 
We selected a semiconducting AGNR and a metallic ZGNR, although we verified that our results are not dependent on such choices. 
The continuous lines represent transmission through the electronic Lévy glass, i.e. in the presence of spin-orbital clusters, whose Rashba interaction is $\lambda=0.07t_0$.  The clusters induce an overall decrease in transmission relative to the pristine ones, along with significant mesoscopic fluctuations, which we will analyze in detail below.

\begin{figure*}[htbp!]
    \centering
    \includegraphics[width=0.80\linewidth]{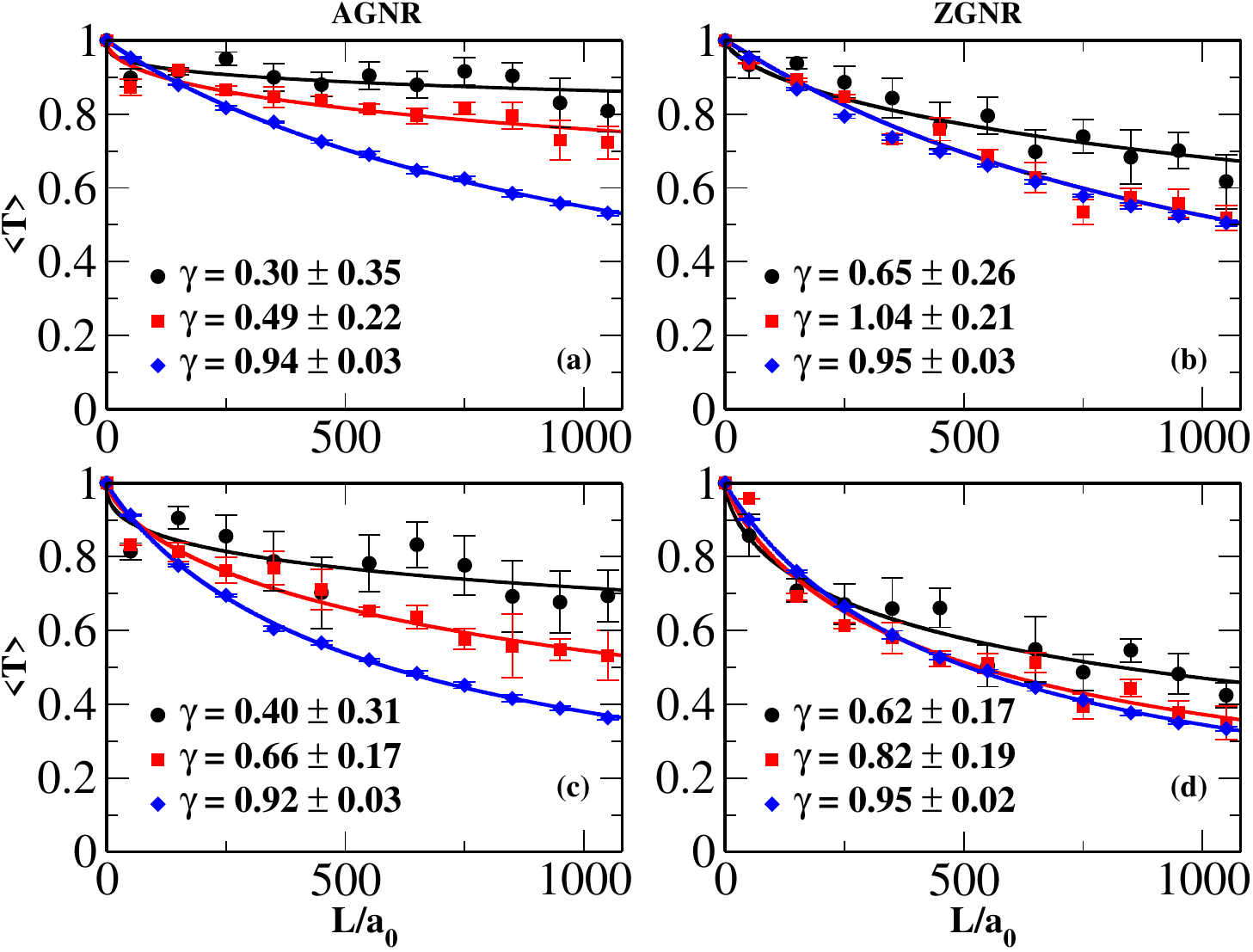}
    \caption{Average transmission through electronic L\'evy glasses as a function of their length $L$, with Rashba interaction (a,b) $\lambda = 0.07t_0$ and (c,d) $\lambda = 0.1t_0$. The left-hand-side panels (a,c) are for AGNR with $N = 2$ (circles), 3 (squares), and 30 (diamonds), while the right-hand-side panels (b,d) are for ZGNR with $N = 3$ (circles), 5 (squares) and 31 (diamonds). The solid lines are fits obtained from Eq. (\ref{eq:alpha}).
}
    \label{fig:T_L}
\end{figure*}

To understand the effect of the spin-orbital clusters on the charge transmission through the electronic Lévy glass, we plot the average charge transmission $\langle T \rangle$ as a function of the device length $L$ for Rashba interactions $\lambda=0.07t_0$ and $0.1t_0$, as shown in Fig. \ref{fig:T_L}. 
We obtained the average charge transmissions (data points), developing an averaging on the charge transmission time series, where the Fermi energy $E$ plays the role of a fictional time. 
The Fermi energy range is defined by the pristine AGNR and ZGNR  transmission steps shown in Fig. \ref{fig:T_E}, which coincide with the number of propagating modes on each ribbon $N$. 
The integer parameter $N$ is directly proportional to the width $W$ and the Fermi wave vector $k_F$ through the relationship $N = k_F W/\pi$. 
Thus, each $N$ value has a different energy range where we numerically calculate a charge transmission time series with 5,000 time steps. 
Finally, we use the latter to obtain the average charge transmission.

Figure \ref{fig:T_L}(a) shows the average transmission $\langle T \rangle$ as a function of $L$ for AGNR with Rashba interaction $\lambda=0.07t_0$ for $N=2$ (circles), $3$ (squares) and 30 (diamonds). 
We can fit the data with Eq. (\ref{eq:alpha}) and obtain the scaling exponent $\gamma$ associated with $N$. 
The exponent characterizes the transport regime, diffusive if $\gamma = 1$ or superdiffusive if $\gamma < 1$. 
For $N=2$, which corresponds to the low Fermi energy regime, the exponent is $\gamma=0.30$. 
It increases to $\gamma=0.94$ for $N=30$, which corresponds to a higher Fermi energy. 
Figure \ref{fig:T_L}(c) shows a similar behavior for AGNR with $\lambda=0.1 t_0$, i.e, when $N=2$ (low energy) the exponent is $\gamma=0.4$, and it increases to $\gamma=0.92$ for $N=30$ (high energy). 
Figures \ref{fig:T_L}(b,d) show $\langle T \rangle$ as a function of $L$ for ZGNR with $\lambda=0.07t_0$ and $0.1 t_0$, respectively, and $N=3$ (circles), $5$ (squares) and 31 (diamonds). 
Similarly to what happens in the AGNR, when $N=3$ we have $\gamma \approx 0.6$, and it increases to 0.95 for $N=31$.

Analysing Fig. \ref{fig:T_L}, we notice that when $N$ is large, i.e. at higher energies, the scaling exponent is close to one, $\gamma \simeq  1$, indicating a diffusive transport regime. 
Conversely, when $N$ is small, i.e. at low energies (also close to Dirac point), the exponent $\gamma \simeq  0.5$, which indicates a superdiffusive transport regime. 
These findings indicate that the presence of spin-orbital clusters can induce Lévy charge transport in graphene nanoribbons. 
Furthermore, the electronic Lévy glass with spin-orbital clusters shows a transition from superdiffusive to diffusive charge transport, as the Fermi energy varies. 
We also stress that those results are independent of whether the electronic Lévy glass is built with an AGNR or a ZGNR.  

\begin{figure}
    \centering
    \includegraphics[width=1.0\linewidth]{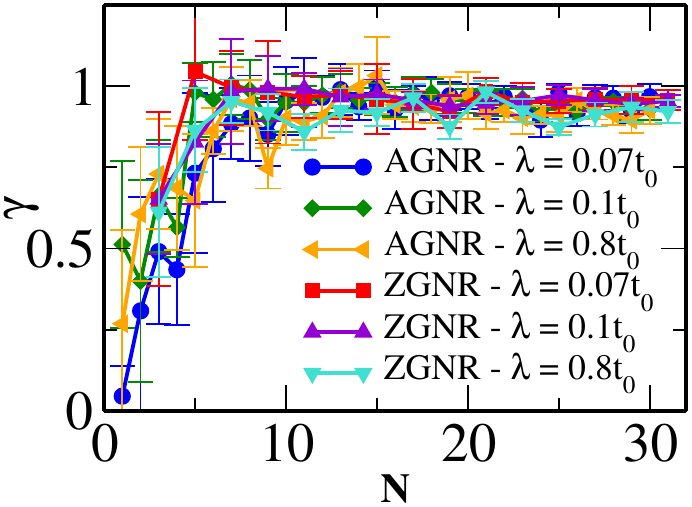}
    \caption{Scaling exponent $\gamma$ as a function of $N$ for AGNR- and ZGNR-based electronic L\'evy glasses in the presence of spin-orbital clusters with Rashba interaction strength $\lambda = 0.07t_0, 0.1t_0$, and $ 0.8t_0$. 
    }
    \label{fig:scaling}
\end{figure}

The superdiffusive-to-diffusive charge transport transition can be further investigated by plotting the scaling exponent as a function of the number of propagating modes. 
Figure \ref{fig:scaling} shows $\gamma$ as a function of $N$ for AGNR and ZGNR electronic L\'evy glasses with Rashba interactions $\lambda = 0.07t_0$, $0.1t_0$, and $0.8t_0$. 
For $N > 10$, the data exhibit stable values around $\gamma = 1$, which means a regular diffusive transport. 
Meanwhile, for $N < 10$, the scaling exponent $\gamma \lesssim 1$, which indicates a change to a superdiffusive transport regime. 
Notably, the results are independent of the Rashba interaction strength and of the type of ribbon edge, which asserts that spin-orbital clusters induce a Lévy-type electronic transport at low energies.

\begin{figure*}
    \centering
    \includegraphics[width=0.7\linewidth]{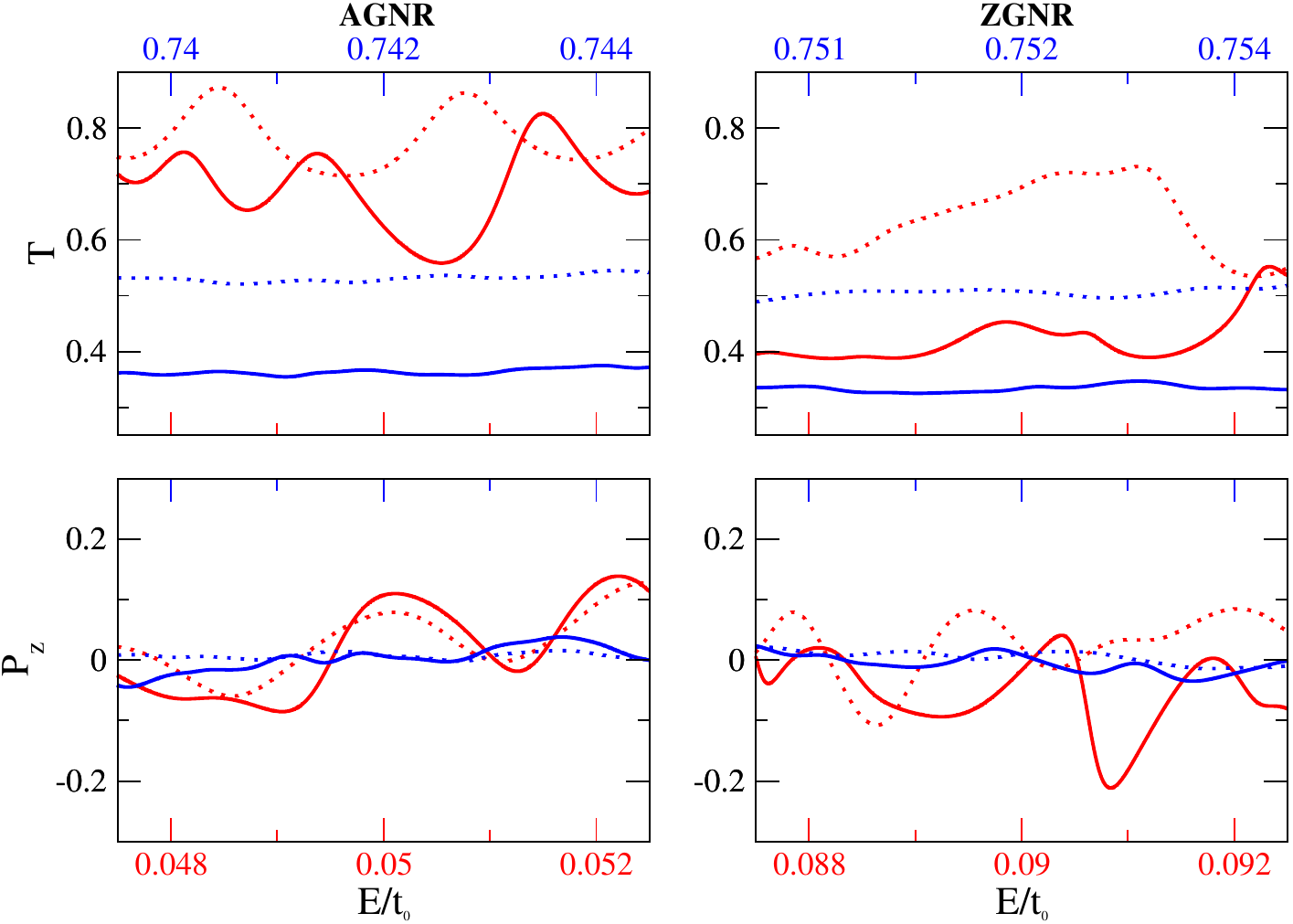}
    \caption{Charge transmission $T$ and spin polarization $P_{z}$ as a function of Fermi energy $E$. The left-hand-side column is for AGNR with $N = 2$ (red lines) and $N = 30$ (blue lines), and the right-hand-side is for ZGNR with $N = 3$ (red lines) and $N = 31$ (blue lines). Dashed and solid lines refer to $\lambda = 0.07t_0$ and $\lambda = 0.1t_0$, respectively.}
    \label{fig:TeP_E}
\end{figure*}

Aside from its charge transport properties, the electronic Lévy glass in the presence of spin-orbital clusters is also expected to exhibit spin polarization transport. 
In order to investigate the effect of the clusters on the spin polarization, we analyze the polarizations defined in Eq. \ref{Pxyz}, as a function of device length, Fermi energy (i.e. number of propagating modes $N$), and Rashba interaction strength. 
As the spin-orbital cluster distribution breaks the transverse reflection symmetry \cite{PhysRevLett.94.246601, Liu_2012}, the spin polarization is not expected to vanish in all three directions: $x$, $y$, and $z$. 

\begin{figure}
    \centering
    \includegraphics[width=1.0\linewidth]{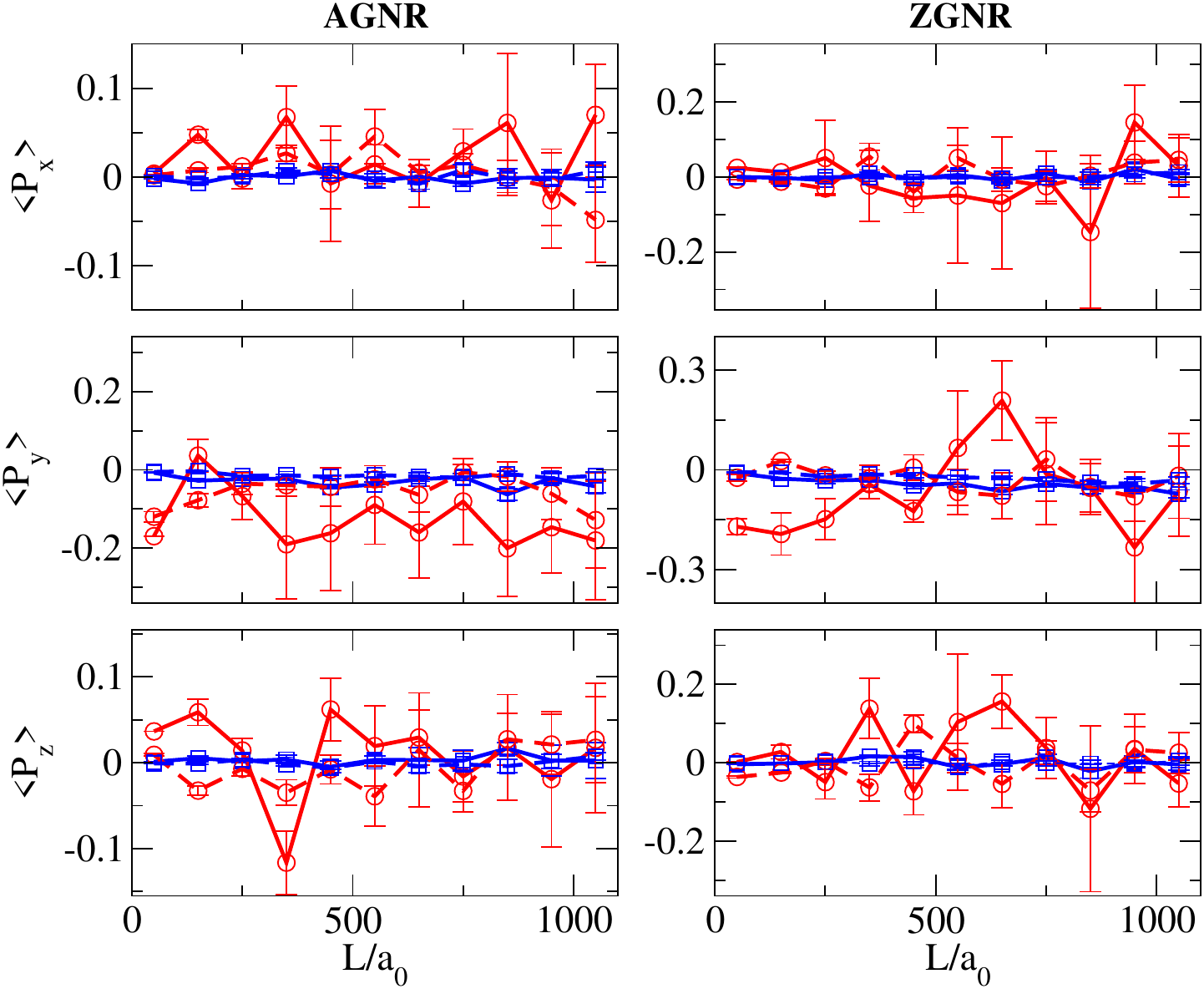}
    \caption{Average spin polarization $\left \langle P_{x,y,z} \right \rangle$ as a function of device length $L$. The left column is for AGNR with $N = 2$ (red symbols) and $N = 30$ (blue symbols), and the right column is for ZGNR with $N = 3$ (red symbols) and $N = 31$ (blue symbols). Dashed and solid lines refer to $\lambda = 0.07t_0$ and $\lambda = 0.1t_0$, respectively, and work as an eye guide.}
    \label{fig:P_L}
\end{figure}

\begin{figure*}
    \centering
    \includegraphics[width=0.8\linewidth]{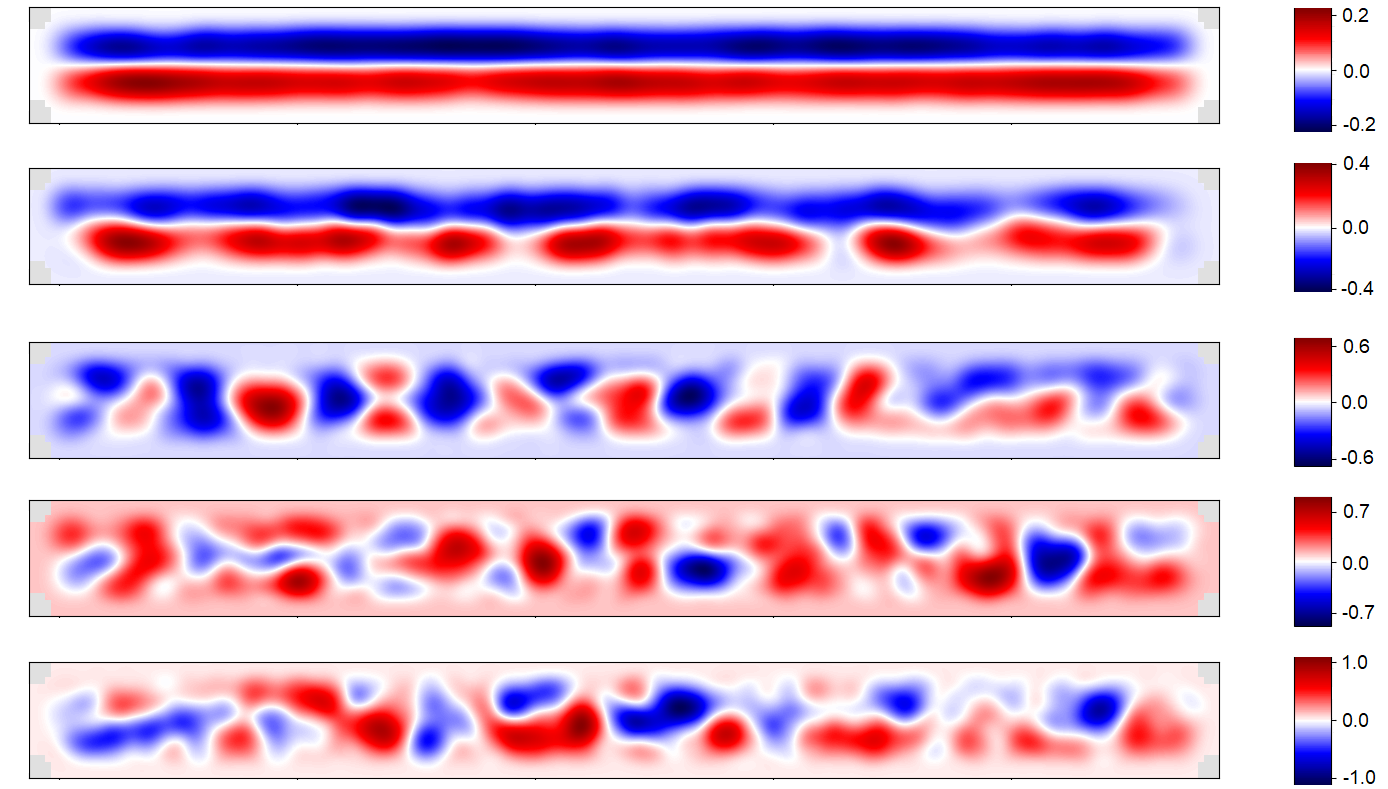}
    \caption{Spin-polarized density of states $\left \langle S_z \right \rangle$ for AGNR electronic L\'evy glass with $\lambda = 0.1t_0$ for $N=1$, 3, 5, 9  and 30 from top to bottom.}
    \label{fig:DOS}
\end{figure*}

Figure \ref{fig:TeP_E} shows the charge transmission $T$ and spin polarization $P_{z}$ as a function of Fermi energy $E$. 
The behavior of $P_{x}$ and $P_{y}$ is qualitatively equivalent to $P_{z}$.
Dashed and solid lines refer to $\lambda = 0.07t_0$ and $\lambda = 0.1t_0$, respectively. 
Panels on the left-hand-side correspond to AGNR with $N = 2$ (red lines, bottom axis) and $N = 30$ (blue lines, top axis), while the right-hand-side ones are for ZGNR with $N = 3$ (red lines, bottom axis) and $N = 31$ (blue lines, top axis). 
The widths of the ribbons are $W_A = 49.5a_0$ and $W_Z = 49.6a_0$ and lengths $L_A = 1050.7a_0$ and $L_Z = 1050.5a_0$. 
It is important to keep in mind that, according to the data in Fig. \ref{fig:scaling}, regular diffusive transport happens for $N>10$, and superdiffusive transport for $N<10$.

As shown in the top panels of Fig. \ref{fig:TeP_E}, the transmission is practically constant in the diffusive transport regime (blue lines), i.e., without significant mesoscopic fluctuations. 
At the same time, the spin polarization is approximately null (down panels), independently of Rashba interaction strength and edge type. 
Thus, the presence of spin-orbital clusters do not induce any spin polarization in the diffusive regime. 
On the other hand, charge transmission shows large mesoscopic fluctuations in the superdiffusive transport regime (red lines), which gives rise to spin polarization with large mesoscopic fluctuations for both edge types, and increase with the Rashba interaction strength.
Thus, the presence of spin-orbital clusters is capable of inducing a spin polarization in the superdiffusive regime only. 

Figure \ref{fig:P_L} shows the average spin polarization $\left \langle P_{x,y,z} \right \rangle$ as a function of device length $L$.
Dashed and solid lines refer to $\lambda = 0.07t_0$ and $\lambda = 0.1t_0$, respectively, and serve as an eye guide. 
Panels on the left-hand-side are for AGNR with $N = 2$ (red symbols) and $N = 30$ (blue symbols), while the right-hand-side panels are for ZGNR with $N = 3$ (red symbols) and $N = 31$ (blue symbols). 
The results of Fig. \ref{fig:P_L} agree with the ones of Fig. \ref{fig:TeP_E}. 
When the transport is in a diffusive regime (blue symbols), the average spin polarization is approximately null, independently of device length, Rashba interaction, and edge type. 
On the other hand, when the transport is in a superdiffusive regime (red symbols), the average spin polarization is significant for all device lengths.

The spin polarization behavior shown in Figs. \ref{fig:TeP_E} and  \ref{fig:P_L} can be rationalized by looking at the spin-polarized density of states. 
Figure \ref{fig:DOS} shows $\left \langle S_z \right \rangle$ for AGNR electronic L\'evy glasses with $\lambda = 0.1t_0$ for $N=1$, 3, 5, 9  and 30 from top to bottom. 
The presence of spin-orbital clusters induce an efficient spin-up and spin-down separation in the superdiffusive regime. 
In practice, this spin segregation means that electrons are submitted to a low magnetoresistivity, which gives rise to high charge transport $T$ and spin polarization $P_z$, as shown in Fig. \ref{fig:TeP_E}. 
Conversely, in the diffusive regime, spin-up and spin-down electrons are not segregated, which originate a large magnetoresistivity, decreasing the charge transport $T$ and suppressesing spin polarization $P_z$.
This feature makes the electronic Lévy glass in the presence of spin-orbital clusters an efficient spin  filter, which can be tuned by its Fermi energy.

\section{Discussion}\label{sec:discussion}

According to Fig. \ref{fig:scaling} the electronic Lévy glass with spin-orbital clusters show a superdiffusive-to-diffusive charge transport transition as the Fermi energy increases. 
This means that for low energies ($N<10$), the charge transport is superdiffusive, while for high energy ($N>10$), it is diffusive. 
We have recently reported a similar transition on an electronic Lévy glass with electrostatic clusters but without SOC in Ref. [\onlinecite{PhysRevB.107.155432}]. 
Thus, we conclude that electrostatic clusters and spin-orbital clusters can independently induce Lévy-like electronic transport in graphene nanoribbons. 

\begin{figure*}
    \centering
    \includegraphics[width=0.47\linewidth]{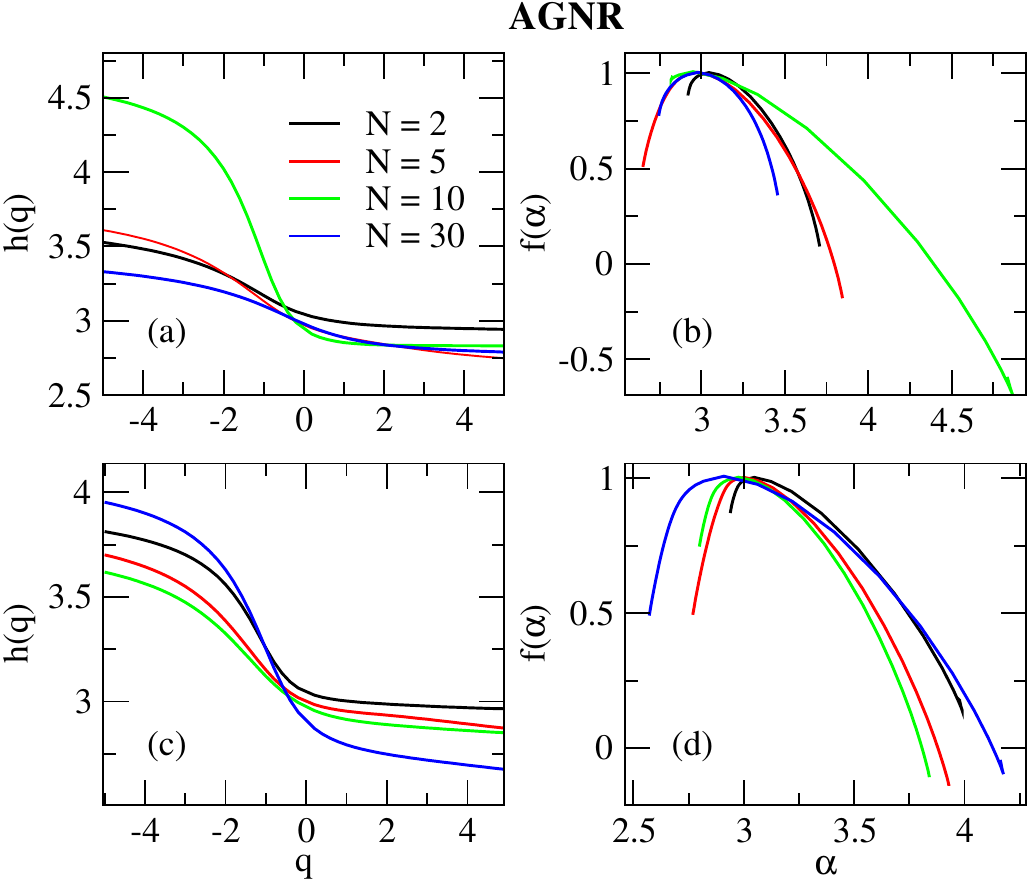}
    \includegraphics[width=0.47\linewidth]{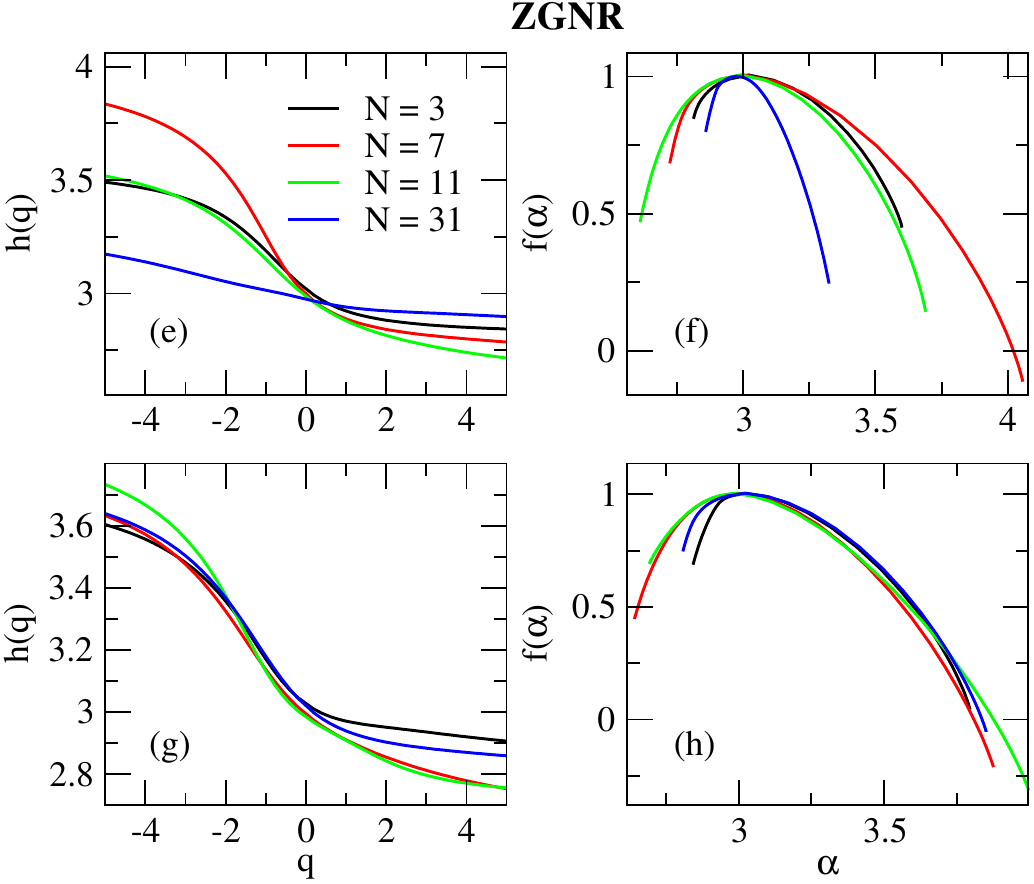}
    \caption{ Multifractal analysis of charge transmission $T$ and spin polarization $P_z$ time series of AGNR (left panels, a-d) and ZGNR  (right panels, e-h) for $\lambda=0.07t_0$ and different $N$. (a,e) Generalized Hurst exponent $h(q)$ as a function of $q$ and (b,f) multifractal singularity spectrum $f(\alpha$) as a function of $\alpha$ obtained from charge transmission time series. (c,g) $h(q)$ and (d,h) $f(\alpha$) obtained from spin polarization $P_z$ time series.}
    \label{fig:Hq_ZGNR_0.07}
\end{figure*}

Furthermore, the electronic Lévy glass with spin-orbital clusters presents spin polarization in the superdiffusive regime, while in the diffusive regime it does not.
In other words, we have high spin polarization efficiency for low Fermi energy (close to Dirac point) and low efficiency for high energy. 
High spin polarization efficiency is related to more significant mesoscopic fluctuations induced by the spin-orbital clusters in the superdiffusive regime in contrast with weak mesoscopic fluctuations in the diffusive regime, as shown in Fig. \ref{fig:TeP_E}.
These features make the electronic Lévy glass a candidate for spintronics applications because its spin polarization can be tuned using its Fermi energy, making it an efficient spin filter as shown in Fig. \ref{fig:DOS}.

As in the case of the electrostatic clusters, the breaking of chiral symmetry is the most compelling explanation for the transport regime transition observed in our calculations \cite{PhysRevB.107.155432}, as well as for the change in spin polarization efficiency. 
The chiral symmetry is preserved at low energies (close to the Dirac point) while it's broken in the high energy limit \cite{PhysRevE.69.056219}, which could induce the transition from the Lévy transport regime to the diffusive one. 
Hence, the observed superdiffusive-to-diffusive transition can be understood as a signature of chiral symmetry breaking. 
In fact, Refs. [\onlinecite{PhysRevLett.111.056801,PhysRevLett.126.206804}] have shown that chiral symmetry breaking can be associated with a non-equilibrium phase transition. 
Thus, a multifractal analysis of charge transport and spin polarization time series is fundamental because it can identify such phase transitions. 
As shown in Ref. [\onlinecite{Zhao_2017}], when we approach a phase transition, there is an increase in multifractality, i.e., an increase in the variation of the generalized Hurst exponent as a function of $q$ and, therefore, an increase in width of $\Delta \alpha$.

Figure \ref{fig:Hq_ZGNR_0.07} shows the multifractal analysis of charge transmission $T$ and spin polarization $P_z$ time series for AGNR (left panels, a-d) and ZGNR  (right panels, e-h) for $\lambda=0.07t_0$ and different $N$ values. 
Figures \ref{fig:Hq_ZGNR_0.07}(a,e) show the generalized Hurst exponent $h(q)$ as a function of $q$ while Figs. \ref{fig:Hq_ZGNR_0.07}(b,f) show the multifractal singularity spectra $f(\alpha$) as functions of $\alpha$ obtained from the charge transmission time series. 
For AGNR, when $N=2$ and 5, $h(\alpha)$ varies with $q$, indicating a multifractal time series in the superdiffusive regime; when $N=30$, the variation decreases indicating a multifractality weakening (tendency to monofractality) in the diffusive regime,  Fig. \ref{fig:Hq_ZGNR_0.07}(a). 
This result is supported by Fig. \ref{fig:Hq_ZGNR_0.07}(b) where $\Delta\alpha$ is large when $N=2$ and 5, and decreases when $N=30$. 
The same analysis is valid for the ZGNR with $N=3$ and $N=31$ in Figs. \ref{fig:Hq_ZGNR_0.07}(e,f). 
This tendency to lose multifractality with an increase in $N$ agrees with recent experimental and theoretical studies \cite{nature,PhysRevE.104.054129}, which developed the multifractal analysis of magnetoconductance time series in graphene ribbons with standard Anderson disorder.
On the other hand, for electronic Lévy glasses when $N=10$ and $N=7$ for AGNR and ZGNR, respectively, we see a significant increase in the amplitude of $h(q)$ and the width of $\Delta\alpha$, which is an indication that we are close to a non-equilibrium phase transition. 
This result supports the hypothesis that the superdiffusive-to-diffusive transition is a signature of chiral symmetry breaking. 

Figures \ref{fig:Hq_ZGNR_0.07}(c,g) show $h(q)$ and Figs. \ref{fig:Hq_ZGNR_0.07}(d,h) show $f(\alpha$) obtained from the spin polarization $P_z$ time series. 
$h(\alpha)$ varied significantly with $q$ independent of $N$ and edge type. 
This is supported by the width of $\Delta\alpha$ that is large. 
This indicates that the spin polarization time series is always multifractal, in contrast with the charge transport time series, which is multifractal for $N<10$ and tends to be monofractal for $N>10$. 
Although the multifractal analysis of the spin polarization time series does not provide information about the breaking of chiral symmetry, it shows that the multifractality of the spin polarization time series is not lost in the diffusive regime, i.e., high energy limit, as opposed to what happens in the case of charge transport. 
This is a notable feature of mesoscopic fluctuations of spin polarization in contrast to mesoscopic fluctuations of charge transport. 
Furthermore, the findings of Fig. \ref{fig:Hq_ZGNR_0.07} are qualitatively equivalent to what we find for $\lambda =0.1t_0$ and spin polarization directions $P_{x,y}$.

Finally, we must discuss the possible origin of the multifractality in the time series of charge transport and spin polarization reported in our results. 
As one might recall, the density of states is sensible to Fermi energy variations. 
When we change the energy, it changes the density of states, which can induce correlations in the time series.  
The correlations are robust for charge transport time series when $N<10$, increasing further as we approach the phase transition, leading to the multifractal time series. 
However, the correlations weaken as $N$ increases, leading to a loss of multifractality in the time series for $N>10$.
On the other hand, the correlations induced by the density of states in the spin polarization time series are always robust, making its time series multifractal for all values of $N$. 
This confirms that spin polarization is more sensible to variations in the density of states, when compared to pure charge transport.

\section{Conclusions} \label{sec:conslusions}

In summary, we proposed an electronic Lévy glass composed of graphene nanoribbons in the presence of  regions with enhanced spin-orbit coupling.  
We considered nanoribbons with armchair and zigzag edges containing circular regions with a tunable Rashba spin-orbit coupling, whose diameter follows a power-law distribution. 
We found that spin-orbital clusters induce a transition from superdiffusive to diffusive charge transport, similar to our recent investigation on nanoribbons with electrostatic clusters \cite{PhysRevB.107.155432}.
We also investigated spin polarization in the spin-orbital Lévy glasses and observed that a finite spin polarization can be found only in the superdiffusive regime.
In contrast, the spin polarization vanishes in the diffusive regime, making the electronic L\'evy glass a useful spintronics device whose electronic transmission and spin polarization can be controlled by its Fermi energy.
Finally, we applied a multifractal analysis to charge transmission and spin polarization and found multifractal transmission time series in the superdiffusive regime, while they tend to be monofractal in the diffusive regime.
In contrast, spin polarization time series are multifractal in both regimes, characterizing a marked difference between mesoscopic fluctuations of charge transport and spin polarization in our electronic L\'evy glass.

\section*{Acknowledgements}

DBF acknowledges a scholarship from Funda\c{c}\~ao de Amparo a Ci\^encia e Tecnologia de Pernambuco (FACEPE, Grant IBPG-0253-1.04/22).
LFCP acknowledges financial support from CAPES (Grant 0041/2022), CNPq (Grants 436859/2018, 313462/2020,  200296/2023-0 and 371610/2023-0 INCT Materials Informatics), FACEPE (Grant APQ-1117-1.05/22), FINEP (Grant 0165/21) and the visiting professors program at Sapienza.
ALRB also acknowledges financial support from CNPq (Grant 309457/2021).

\bibliographystyle{apsrev4-1}
\bibliography{ref.bib}

\begin{thebibliography}{47}%
\makeatletter
\providecommand \@ifxundefined [1]{%
 \@ifx{#1\undefined}
}%
\providecommand \@ifnum [1]{%
 \ifnum #1\expandafter \@firstoftwo
 \else \expandafter \@secondoftwo
 \fi
}%
\providecommand \@ifx [1]{%
 \ifx #1\expandafter \@firstoftwo
 \else \expandafter \@secondoftwo
 \fi
}%
\providecommand \natexlab [1]{#1}%
\providecommand \enquote  [1]{``#1''}%
\providecommand \bibnamefont  [1]{#1}%
\providecommand \bibfnamefont [1]{#1}%
\providecommand \citenamefont [1]{#1}%
\providecommand \href@noop [0]{\@secondoftwo}%
\providecommand \href [0]{\begingroup \@sanitize@url \@href}%
\providecommand \@href[1]{\@@startlink{#1}\@@href}%
\providecommand \@@href[1]{\endgroup#1\@@endlink}%
\providecommand \@sanitize@url [0]{\catcode `\\12\catcode `\$12\catcode
  `\&12\catcode `\#12\catcode `\^12\catcode `\_12\catcode `\%12\relax}%
\providecommand \@@startlink[1]{}%
\providecommand \@@endlink[0]{}%
\providecommand \url  [0]{\begingroup\@sanitize@url \@url }%
\providecommand \@url [1]{\endgroup\@href {#1}{\urlprefix }}%
\providecommand \urlprefix  [0]{URL }%
\providecommand \Eprint [0]{\href }%
\providecommand \doibase [0]{http://dx.doi.org/}%
\providecommand \selectlanguage [0]{\@gobble}%
\providecommand \bibinfo  [0]{\@secondoftwo}%
\providecommand \bibfield  [0]{\@secondoftwo}%
\providecommand \translation [1]{[#1]}%
\providecommand \BibitemOpen [0]{}%
\providecommand \bibitemStop [0]{}%
\providecommand \bibitemNoStop [0]{.\EOS\space}%
\providecommand \EOS [0]{\spacefactor3000\relax}%
\providecommand \BibitemShut  [1]{\csname bibitem#1\endcsname}%
\let\auto@bib@innerbib\@empty
\bibitem [{\citenamefont {Editorial}(2024)}]{ref1}%
  \BibitemOpen
  \bibfield  {author} {\bibinfo {author} {\bibnamefont {Editorial}},\ }\href
  {\doibase 10.1038/s41567-023-02381-0} {\bibfield  {journal} {\bibinfo
  {journal} {Nature Physics}\ }\textbf {\bibinfo {volume} {20}} (\bibinfo
  {year} {2024}),\ 10.1038/s41567-023-02381-0}\BibitemShut {NoStop}%
\bibitem [{\citenamefont {Avsar}\ \emph {et~al.}(2020)\citenamefont {Avsar},
  \citenamefont {Ochoa}, \citenamefont {Guinea}, \citenamefont {\"Ozyilmaz},
  \citenamefont {van Wees},\ and\ \citenamefont
  {Vera-Marun}}]{RevModPhys.92.021003}%
  \BibitemOpen
  \bibfield  {author} {\bibinfo {author} {\bibfnamefont {A.}~\bibnamefont
  {Avsar}}, \bibinfo {author} {\bibfnamefont {H.}~\bibnamefont {Ochoa}},
  \bibinfo {author} {\bibfnamefont {F.}~\bibnamefont {Guinea}}, \bibinfo
  {author} {\bibfnamefont {B.}~\bibnamefont {\"Ozyilmaz}}, \bibinfo {author}
  {\bibfnamefont {B.~J.}\ \bibnamefont {van Wees}}, \ and\ \bibinfo {author}
  {\bibfnamefont {I.~J.}\ \bibnamefont {Vera-Marun}},\ }\href {\doibase
  10.1103/RevModPhys.92.021003} {\bibfield  {journal} {\bibinfo  {journal}
  {Rev. Mod. Phys.}\ }\textbf {\bibinfo {volume} {92}},\ \bibinfo {pages}
  {021003} (\bibinfo {year} {2020})}\BibitemShut {NoStop}%
\bibitem [{\citenamefont {Sierra}\ \emph {et~al.}(2021)\citenamefont {Sierra},
  \citenamefont {Fabian}, \citenamefont {Kawakami}, \citenamefont {Roche},\
  and\ \citenamefont {Valenzuela}}]{Sierra_2021}%
  \BibitemOpen
  \bibfield  {author} {\bibinfo {author} {\bibfnamefont {J.~F.}\ \bibnamefont
  {Sierra}}, \bibinfo {author} {\bibfnamefont {J.}~\bibnamefont {Fabian}},
  \bibinfo {author} {\bibfnamefont {R.~K.}\ \bibnamefont {Kawakami}}, \bibinfo
  {author} {\bibfnamefont {S.}~\bibnamefont {Roche}}, \ and\ \bibinfo {author}
  {\bibfnamefont {S.~O.}\ \bibnamefont {Valenzuela}},\ }\href {\doibase
  10.1038/s41565-021-00936-x} {\bibfield  {journal} {\bibinfo  {journal}
  {Nature Nanotechnology}\ }\textbf {\bibinfo {volume} {16}},\ \bibinfo {pages}
  {856–868} (\bibinfo {year} {2021})}\BibitemShut {NoStop}%
\bibitem [{\citenamefont {Perkins}\ and\ \citenamefont
  {Ferreira}(2024)}]{Perkins_2024}%
  \BibitemOpen
  \bibfield  {author} {\bibinfo {author} {\bibfnamefont {D.~T.}\ \bibnamefont
  {Perkins}}\ and\ \bibinfo {author} {\bibfnamefont {A.}~\bibnamefont
  {Ferreira}},\ }\enquote {\bibinfo {title} {Spintronics in 2d graphene-based
  van der waals heterostructures},}\ in\ \href {\doibase
  10.1016/b978-0-323-90800-9.00203-1} {\emph {\bibinfo {booktitle}
  {Encyclopedia of Condensed Matter Physics}}}\ (\bibinfo  {publisher}
  {Elsevier},\ \bibinfo {year} {2024})\ p.\ \bibinfo {pages}
  {205–222}\BibitemShut {NoStop}%
\bibitem [{\citenamefont {Alcón}\ \emph {et~al.}(2024)\citenamefont {Alcón},
  \citenamefont {Cummings},\ and\ \citenamefont {Roche}}]{D3NH00416C}%
  \BibitemOpen
  \bibfield  {author} {\bibinfo {author} {\bibfnamefont {I.}~\bibnamefont
  {Alcón}}, \bibinfo {author} {\bibfnamefont {A.~W.}\ \bibnamefont
  {Cummings}}, \ and\ \bibinfo {author} {\bibfnamefont {S.}~\bibnamefont
  {Roche}},\ }\href {\doibase 10.1039/D3NH00416C} {\bibfield  {journal}
  {\bibinfo  {journal} {Nanoscale Horiz.}\ }\textbf {\bibinfo {volume} {9}},\
  \bibinfo {pages} {407} (\bibinfo {year} {2024})}\BibitemShut {NoStop}%
\bibitem [{\citenamefont {Bihlmayer}\ \emph {et~al.}(2022)\citenamefont
  {Bihlmayer}, \citenamefont {N{\"o}el}, \citenamefont {Vyalikh}, \citenamefont
  {Chulkov},\ and\ \citenamefont {Manchon}}]{Bihlmayer2022RashbalikePI}%
  \BibitemOpen
  \bibfield  {author} {\bibinfo {author} {\bibfnamefont {G.}~\bibnamefont
  {Bihlmayer}}, \bibinfo {author} {\bibfnamefont {P.}~\bibnamefont {N{\"o}el}},
  \bibinfo {author} {\bibfnamefont {D.~V.}\ \bibnamefont {Vyalikh}}, \bibinfo
  {author} {\bibfnamefont {E.~V.}\ \bibnamefont {Chulkov}}, \ and\ \bibinfo
  {author} {\bibfnamefont {A.}~\bibnamefont {Manchon}},\ }\href
  {https://api.semanticscholar.org/CorpusID:251831291} {\bibfield  {journal}
  {\bibinfo  {journal} {Nature Reviews Physics}\ }\textbf {\bibinfo {volume}
  {4}},\ \bibinfo {pages} {642 } (\bibinfo {year} {2022})}\BibitemShut
  {NoStop}%
\bibitem [{\citenamefont {Nitta}(2023)}]{JunsakuNitta2023230102}%
  \BibitemOpen
  \bibfield  {author} {\bibinfo {author} {\bibfnamefont {J.}~\bibnamefont
  {Nitta}},\ }\href {\doibase 10.11470/jsaprev.230102} {\bibfield  {journal}
  {\bibinfo  {journal} {JSAP Review}\ }\textbf {\bibinfo {volume} {2023}},\
  \bibinfo {pages} {230102} (\bibinfo {year} {2023})}\BibitemShut {NoStop}%
\bibitem [{\citenamefont {Mendes}\ \emph {et~al.}(2019)\citenamefont {Mendes},
  \citenamefont {Alves~Santos}, \citenamefont {Chagas}, \citenamefont
  {Magalh\~aes Paniago}, \citenamefont {Mori}, \citenamefont {Holanda},
  \citenamefont {Meireles}, \citenamefont {Lacerda}, \citenamefont {Azevedo},\
  and\ \citenamefont {Rezende}}]{PhysRevB.99.214446}%
  \BibitemOpen
  \bibfield  {author} {\bibinfo {author} {\bibfnamefont {J.~B.~S.}\
  \bibnamefont {Mendes}}, \bibinfo {author} {\bibfnamefont {O.}~\bibnamefont
  {Alves~Santos}}, \bibinfo {author} {\bibfnamefont {T.}~\bibnamefont
  {Chagas}}, \bibinfo {author} {\bibfnamefont {R.}~\bibnamefont {Magalh\~aes
  Paniago}}, \bibinfo {author} {\bibfnamefont {T.~J.~A.}\ \bibnamefont {Mori}},
  \bibinfo {author} {\bibfnamefont {J.}~\bibnamefont {Holanda}}, \bibinfo
  {author} {\bibfnamefont {L.~M.}\ \bibnamefont {Meireles}}, \bibinfo {author}
  {\bibfnamefont {R.~G.}\ \bibnamefont {Lacerda}}, \bibinfo {author}
  {\bibfnamefont {A.}~\bibnamefont {Azevedo}}, \ and\ \bibinfo {author}
  {\bibfnamefont {S.~M.}\ \bibnamefont {Rezende}},\ }\href {\doibase
  10.1103/PhysRevB.99.214446} {\bibfield  {journal} {\bibinfo  {journal} {Phys.
  Rev. B}\ }\textbf {\bibinfo {volume} {99}},\ \bibinfo {pages} {214446}
  (\bibinfo {year} {2019})}\BibitemShut {NoStop}%
\bibitem [{\citenamefont {Liu}\ \emph {et~al.}(2012)\citenamefont {Liu},
  \citenamefont {Chan},\ and\ \citenamefont {Wang}}]{Liu_2012}%
  \BibitemOpen
  \bibfield  {author} {\bibinfo {author} {\bibfnamefont {J.-F.}\ \bibnamefont
  {Liu}}, \bibinfo {author} {\bibfnamefont {K.~S.}\ \bibnamefont {Chan}}, \
  and\ \bibinfo {author} {\bibfnamefont {J.}~\bibnamefont {Wang}},\ }\href
  {\doibase 10.1088/0957-4484/23/9/095201} {\bibfield  {journal} {\bibinfo
  {journal} {Nanotechnology}\ }\textbf {\bibinfo {volume} {23}},\ \bibinfo
  {pages} {095201} (\bibinfo {year} {2012})}\BibitemShut {NoStop}%
\bibitem [{\citenamefont {Zhang}\ \emph {et~al.}(2014)\citenamefont {Zhang},
  \citenamefont {Chan},\ and\ \citenamefont {Lin}}]{Zhang_2014}%
  \BibitemOpen
  \bibfield  {author} {\bibinfo {author} {\bibfnamefont {Q.}~\bibnamefont
  {Zhang}}, \bibinfo {author} {\bibfnamefont {K.~S.}\ \bibnamefont {Chan}}, \
  and\ \bibinfo {author} {\bibfnamefont {Z.}~\bibnamefont {Lin}},\ }\href
  {\doibase 10.1088/0022-3727/47/43/435302} {\bibfield  {journal} {\bibinfo
  {journal} {Journal of Physics D: Applied Physics}\ }\textbf {\bibinfo
  {volume} {47}},\ \bibinfo {pages} {435302} (\bibinfo {year}
  {2014})}\BibitemShut {NoStop}%
\bibitem [{\citenamefont {Santos}\ \emph {et~al.}(2020)\citenamefont {Santos},
  \citenamefont {Latgé}, \citenamefont {Brey},\ and\ \citenamefont
  {Chico}}]{SANTOS20201}%
  \BibitemOpen
  \bibfield  {author} {\bibinfo {author} {\bibfnamefont {H.}~\bibnamefont
  {Santos}}, \bibinfo {author} {\bibfnamefont {A.}~\bibnamefont {Latgé}},
  \bibinfo {author} {\bibfnamefont {L.}~\bibnamefont {Brey}}, \ and\ \bibinfo
  {author} {\bibfnamefont {L.}~\bibnamefont {Chico}},\ }\href {\doibase
  https://doi.org/10.1016/j.carbon.2020.05.054} {\bibfield  {journal} {\bibinfo
   {journal} {Carbon}\ }\textbf {\bibinfo {volume} {168}},\ \bibinfo {pages}
  {1} (\bibinfo {year} {2020})}\BibitemShut {NoStop}%
\bibitem [{\citenamefont {Belayadi}\ and\ \citenamefont
  {Vasilopoulos}(2023)}]{Belayadi_2023}%
  \BibitemOpen
  \bibfield  {author} {\bibinfo {author} {\bibfnamefont {A.}~\bibnamefont
  {Belayadi}}\ and\ \bibinfo {author} {\bibfnamefont {P.}~\bibnamefont
  {Vasilopoulos}},\ }\href {\doibase 10.1088/1361-6528/acd8c1} {\bibfield
  {journal} {\bibinfo  {journal} {Nanotechnology}\ }\textbf {\bibinfo {volume}
  {34}},\ \bibinfo {pages} {365706} (\bibinfo {year} {2023})}\BibitemShut
  {NoStop}%
\bibitem [{\citenamefont {Zhang}\ \emph {et~al.}(2017)\citenamefont {Zhang},
  \citenamefont {Chan},\ and\ \citenamefont {Li}}]{C6CP06972J}%
  \BibitemOpen
  \bibfield  {author} {\bibinfo {author} {\bibfnamefont {Q.}~\bibnamefont
  {Zhang}}, \bibinfo {author} {\bibfnamefont {K.~S.}\ \bibnamefont {Chan}}, \
  and\ \bibinfo {author} {\bibfnamefont {J.}~\bibnamefont {Li}},\ }\href
  {\doibase 10.1039/C6CP06972J} {\bibfield  {journal} {\bibinfo  {journal}
  {Phys. Chem. Chem. Phys.}\ }\textbf {\bibinfo {volume} {19}},\ \bibinfo
  {pages} {6871} (\bibinfo {year} {2017})}\BibitemShut {NoStop}%
\bibitem [{\citenamefont {Ramos}\ \emph {et~al.}(2018)\citenamefont {Ramos},
  \citenamefont {Vasconcelos},\ and\ \citenamefont
  {Barbosa}}]{10.1063/1.5010973}%
  \BibitemOpen
  \bibfield  {author} {\bibinfo {author} {\bibfnamefont {J.~G. G.~S.}\
  \bibnamefont {Ramos}}, \bibinfo {author} {\bibfnamefont {T.~C.}\ \bibnamefont
  {Vasconcelos}}, \ and\ \bibinfo {author} {\bibfnamefont {A.~L.~R.}\
  \bibnamefont {Barbosa}},\ }\href {\doibase 10.1063/1.5010973} {\bibfield
  {journal} {\bibinfo  {journal} {Journal of Applied Physics}\ }\textbf
  {\bibinfo {volume} {123}},\ \bibinfo {pages} {034304} (\bibinfo {year}
  {2018})},\ \Eprint
  {http://arxiv.org/abs/https://pubs.aip.org/aip/jap/article-pdf/doi/10.1063/1.5010973/15204912/034304\_1\_online.pdf}
  {https://pubs.aip.org/aip/jap/article-pdf/doi/10.1063/1.5010973/15204912/034304\_1\_online.pdf}
  \BibitemShut {NoStop}%
\bibitem [{\citenamefont {da~Silva}\ \emph {et~al.}(2022)\citenamefont
  {da~Silva}, \citenamefont {Santana}, \citenamefont {Ramos},\ and\
  \citenamefont {Barbosa}}]{10.1063/5.0107212}%
  \BibitemOpen
  \bibfield  {author} {\bibinfo {author} {\bibfnamefont {J.~M.}\ \bibnamefont
  {da~Silva}}, \bibinfo {author} {\bibfnamefont {F.~A.~F.}\ \bibnamefont
  {Santana}}, \bibinfo {author} {\bibfnamefont {J.~G. G.~S.}\ \bibnamefont
  {Ramos}}, \ and\ \bibinfo {author} {\bibfnamefont {A.~L.~R.}\ \bibnamefont
  {Barbosa}},\ }\href {\doibase 10.1063/5.0107212} {\bibfield  {journal}
  {\bibinfo  {journal} {Journal of Applied Physics}\ }\textbf {\bibinfo
  {volume} {132}},\ \bibinfo {pages} {183901} (\bibinfo {year} {2022})},\
  \Eprint
  {http://arxiv.org/abs/https://pubs.aip.org/aip/jap/article-pdf/doi/10.1063/5.0107212/16516946/183901\_1\_online.pdf}
  {https://pubs.aip.org/aip/jap/article-pdf/doi/10.1063/5.0107212/16516946/183901\_1\_online.pdf}
  \BibitemShut {NoStop}%
\bibitem [{\citenamefont {Oliver}\ and\ \citenamefont
  {Rappoport}(2016)}]{PhysRevB.94.045432}%
  \BibitemOpen
  \bibfield  {author} {\bibinfo {author} {\bibfnamefont {D.}~\bibnamefont
  {Oliver}}\ and\ \bibinfo {author} {\bibfnamefont {T.~G.}\ \bibnamefont
  {Rappoport}},\ }\href {\doibase 10.1103/PhysRevB.94.045432} {\bibfield
  {journal} {\bibinfo  {journal} {Phys. Rev. B}\ }\textbf {\bibinfo {volume}
  {94}},\ \bibinfo {pages} {045432} (\bibinfo {year} {2016})}\BibitemShut
  {NoStop}%
\bibitem [{\citenamefont {Brede}\ \emph {et~al.}(2023)\citenamefont {Brede},
  \citenamefont {Merino-Díez}, \citenamefont {Berdonces-Layunta},
  \citenamefont {Sanz}, \citenamefont {Domínguez-Celorrio}, \citenamefont
  {Lobo-Checa}, \citenamefont {Vilas-Varela}, \citenamefont {Peña},
  \citenamefont {Frederiksen}, \citenamefont {Pascual}, \citenamefont
  {de~Oteyza},\ and\ \citenamefont {Serrate}}]{Brede_2023}%
  \BibitemOpen
  \bibfield  {author} {\bibinfo {author} {\bibfnamefont {J.}~\bibnamefont
  {Brede}}, \bibinfo {author} {\bibfnamefont {N.}~\bibnamefont {Merino-Díez}},
  \bibinfo {author} {\bibfnamefont {A.}~\bibnamefont {Berdonces-Layunta}},
  \bibinfo {author} {\bibfnamefont {S.}~\bibnamefont {Sanz}}, \bibinfo {author}
  {\bibfnamefont {A.}~\bibnamefont {Domínguez-Celorrio}}, \bibinfo {author}
  {\bibfnamefont {J.}~\bibnamefont {Lobo-Checa}}, \bibinfo {author}
  {\bibfnamefont {M.}~\bibnamefont {Vilas-Varela}}, \bibinfo {author}
  {\bibfnamefont {D.}~\bibnamefont {Peña}}, \bibinfo {author} {\bibfnamefont
  {T.}~\bibnamefont {Frederiksen}}, \bibinfo {author} {\bibfnamefont {J.~I.}\
  \bibnamefont {Pascual}}, \bibinfo {author} {\bibfnamefont {D.~G.}\
  \bibnamefont {de~Oteyza}}, \ and\ \bibinfo {author} {\bibfnamefont
  {D.}~\bibnamefont {Serrate}},\ }\href {\doibase 10.1038/s41467-023-42436-7}
  {\bibfield  {journal} {\bibinfo  {journal} {Nature Communications}\ }\textbf
  {\bibinfo {volume} {14}} (\bibinfo {year} {2023}),\
  10.1038/s41467-023-42436-7}\BibitemShut {NoStop}%
\bibitem [{\citenamefont {Ying}\ and\ \citenamefont
  {Lai}(2016)}]{PhysRevB.93.085408}%
  \BibitemOpen
  \bibfield  {author} {\bibinfo {author} {\bibfnamefont {L.}~\bibnamefont
  {Ying}}\ and\ \bibinfo {author} {\bibfnamefont {Y.-C.}\ \bibnamefont {Lai}},\
  }\href {\doibase 10.1103/PhysRevB.93.085408} {\bibfield  {journal} {\bibinfo
  {journal} {Phys. Rev. B}\ }\textbf {\bibinfo {volume} {93}},\ \bibinfo
  {pages} {085408} (\bibinfo {year} {2016})}\BibitemShut {NoStop}%
\bibitem [{\citenamefont {Park}\ \emph {et~al.}(2020)\citenamefont {Park},
  \citenamefont {Oh}, \citenamefont {Jin}, \citenamefont {Jo}, \citenamefont
  {Choe}, \citenamefont {Yun}, \citenamefont {Lee}, \citenamefont {Lee},
  \citenamefont {Kwon}, \citenamefont {Jin}, \citenamefont {Chung},\ and\
  \citenamefont {Yoo}}]{Park2020ObservationOS}%
  \BibitemOpen
  \bibfield  {author} {\bibinfo {author} {\bibfnamefont {J.}~\bibnamefont
  {Park}}, \bibinfo {author} {\bibfnamefont {I.}~\bibnamefont {Oh}}, \bibinfo
  {author} {\bibfnamefont {M.-J.}\ \bibnamefont {Jin}}, \bibinfo {author}
  {\bibfnamefont {J.}~\bibnamefont {Jo}}, \bibinfo {author} {\bibfnamefont
  {D.}~\bibnamefont {Choe}}, \bibinfo {author} {\bibfnamefont {H.~D.}\
  \bibnamefont {Yun}}, \bibinfo {author} {\bibfnamefont {S.~W.}\ \bibnamefont
  {Lee}}, \bibinfo {author} {\bibfnamefont {Z.}~\bibnamefont {Lee}}, \bibinfo
  {author} {\bibfnamefont {S.}~\bibnamefont {Kwon}}, \bibinfo {author}
  {\bibfnamefont {H.}~\bibnamefont {Jin}}, \bibinfo {author} {\bibfnamefont
  {S.~B.}\ \bibnamefont {Chung}}, \ and\ \bibinfo {author} {\bibfnamefont
  {J.}~\bibnamefont {Yoo}},\ }\href
  {https://api.semanticscholar.org/CorpusID:212718661} {\bibfield  {journal}
  {\bibinfo  {journal} {Scientific Reports}\ }\textbf {\bibinfo {volume} {10}}
  (\bibinfo {year} {2020})}\BibitemShut {NoStop}%
\bibitem [{\citenamefont {Lu}\ and\ \citenamefont {Sun}(2023)}]{Lu_2023}%
  \BibitemOpen
  \bibfield  {author} {\bibinfo {author} {\bibfnamefont {W.-T.}\ \bibnamefont
  {Lu}}\ and\ \bibinfo {author} {\bibfnamefont {Q.-F.}\ \bibnamefont {Sun}},\
  }\href {\doibase 10.1088/1367-2630/accb06} {\bibfield  {journal} {\bibinfo
  {journal} {New Journal of Physics}\ }\textbf {\bibinfo {volume} {25}},\
  \bibinfo {pages} {043018} (\bibinfo {year} {2023})}\BibitemShut {NoStop}%
\bibitem [{\citenamefont {Varykhalov}\ \emph {et~al.}(2008)\citenamefont
  {Varykhalov}, \citenamefont {S\'anchez-Barriga}, \citenamefont {Shikin},
  \citenamefont {Biswas}, \citenamefont {Vescovo}, \citenamefont {Rybkin},
  \citenamefont {Marchenko},\ and\ \citenamefont
  {Rader}}]{PhysRevLett.101.157601}%
  \BibitemOpen
  \bibfield  {author} {\bibinfo {author} {\bibfnamefont {A.}~\bibnamefont
  {Varykhalov}}, \bibinfo {author} {\bibfnamefont {J.}~\bibnamefont
  {S\'anchez-Barriga}}, \bibinfo {author} {\bibfnamefont {A.~M.}\ \bibnamefont
  {Shikin}}, \bibinfo {author} {\bibfnamefont {C.}~\bibnamefont {Biswas}},
  \bibinfo {author} {\bibfnamefont {E.}~\bibnamefont {Vescovo}}, \bibinfo
  {author} {\bibfnamefont {A.}~\bibnamefont {Rybkin}}, \bibinfo {author}
  {\bibfnamefont {D.}~\bibnamefont {Marchenko}}, \ and\ \bibinfo {author}
  {\bibfnamefont {O.}~\bibnamefont {Rader}},\ }\href {\doibase
  10.1103/PhysRevLett.101.157601} {\bibfield  {journal} {\bibinfo  {journal}
  {Phys. Rev. Lett.}\ }\textbf {\bibinfo {volume} {101}},\ \bibinfo {pages}
  {157601} (\bibinfo {year} {2008})}\BibitemShut {NoStop}%
\bibitem [{\citenamefont {Wang}\ \emph {et~al.}(2015)\citenamefont {Wang},
  \citenamefont {Ki}, \citenamefont {Chen}, \citenamefont {Berger},
  \citenamefont {MacDonald},\ and\ \citenamefont {Morpurgo}}]{Wang_2015}%
  \BibitemOpen
  \bibfield  {author} {\bibinfo {author} {\bibfnamefont {Z.}~\bibnamefont
  {Wang}}, \bibinfo {author} {\bibfnamefont {D.}~\bibnamefont {Ki}}, \bibinfo
  {author} {\bibfnamefont {H.}~\bibnamefont {Chen}}, \bibinfo {author}
  {\bibfnamefont {H.}~\bibnamefont {Berger}}, \bibinfo {author} {\bibfnamefont
  {A.~H.}\ \bibnamefont {MacDonald}}, \ and\ \bibinfo {author} {\bibfnamefont
  {A.~F.}\ \bibnamefont {Morpurgo}},\ }\href {\doibase 10.1038/ncomms9339}
  {\bibfield  {journal} {\bibinfo  {journal} {Nature Communications}\ }\textbf
  {\bibinfo {volume} {6}} (\bibinfo {year} {2015}),\
  10.1038/ncomms9339}\BibitemShut {NoStop}%
\bibitem [{\citenamefont {Marchenko}\ \emph {et~al.}(2012)\citenamefont
  {Marchenko}, \citenamefont {Varykhalov}, \citenamefont {Scholz},
  \citenamefont {Bihlmayer}, \citenamefont {Rashba}, \citenamefont {Rybkin},
  \citenamefont {Shikin},\ and\ \citenamefont {Rader}}]{Marchenko2012GiantRS}%
  \BibitemOpen
  \bibfield  {author} {\bibinfo {author} {\bibfnamefont {D.}~\bibnamefont
  {Marchenko}}, \bibinfo {author} {\bibfnamefont {A.}~\bibnamefont
  {Varykhalov}}, \bibinfo {author} {\bibfnamefont {M.~R.}\ \bibnamefont
  {Scholz}}, \bibinfo {author} {\bibfnamefont {G.}~\bibnamefont {Bihlmayer}},
  \bibinfo {author} {\bibfnamefont {E.~I.}\ \bibnamefont {Rashba}}, \bibinfo
  {author} {\bibfnamefont {A.~G.}\ \bibnamefont {Rybkin}}, \bibinfo {author}
  {\bibfnamefont {A.~M.}\ \bibnamefont {Shikin}}, \ and\ \bibinfo {author}
  {\bibfnamefont {O.}~\bibnamefont {Rader}},\ }\href
  {https://api.semanticscholar.org/CorpusID:205314527} {\bibfield  {journal}
  {\bibinfo  {journal} {Nature Communications}\ }\textbf {\bibinfo {volume}
  {3}} (\bibinfo {year} {2012})}\BibitemShut {NoStop}%
\bibitem [{\citenamefont {Caridad}\ \emph {et~al.}(2016)\citenamefont
  {Caridad}, \citenamefont {Connaughton}, \citenamefont {Ott}, \citenamefont
  {Weber},\ and\ \citenamefont {Krsti\'c}}]{Caridad2016}%
  \BibitemOpen
  \bibfield  {author} {\bibinfo {author} {\bibfnamefont {J.~M.}\ \bibnamefont
  {Caridad}}, \bibinfo {author} {\bibfnamefont {S.}~\bibnamefont
  {Connaughton}}, \bibinfo {author} {\bibfnamefont {C.}~\bibnamefont {Ott}},
  \bibinfo {author} {\bibfnamefont {H.~B.}\ \bibnamefont {Weber}}, \ and\
  \bibinfo {author} {\bibfnamefont {V.}~\bibnamefont {Krsti\'c}},\ }\href
  {\doibase doi.org/10.1038/ncomms12894} {\bibfield  {journal} {\bibinfo
  {journal} {Nature Communications}\ }\textbf {\bibinfo {volume} {7}},\
  \bibinfo {pages} {12894} (\bibinfo {year} {2016})}\BibitemShut {NoStop}%
\bibitem [{\citenamefont {Heinisch}\ \emph {et~al.}(2013)\citenamefont
  {Heinisch}, \citenamefont {Bronold},\ and\ \citenamefont
  {Fehske}}]{PhysRevB.87.155409}%
  \BibitemOpen
  \bibfield  {author} {\bibinfo {author} {\bibfnamefont {R.~L.}\ \bibnamefont
  {Heinisch}}, \bibinfo {author} {\bibfnamefont {F.~X.}\ \bibnamefont
  {Bronold}}, \ and\ \bibinfo {author} {\bibfnamefont {H.}~\bibnamefont
  {Fehske}},\ }\href {\doibase 10.1103/PhysRevB.87.155409} {\bibfield
  {journal} {\bibinfo  {journal} {Phys. Rev. B}\ }\textbf {\bibinfo {volume}
  {87}},\ \bibinfo {pages} {155409} (\bibinfo {year} {2013})}\BibitemShut
  {NoStop}%
\bibitem [{\citenamefont {Fehske}\ \emph {et~al.}(2015)\citenamefont {Fehske},
  \citenamefont {Hager},\ and\ \citenamefont
  {Pieper}}]{https://doi.org/10.1002/pssb.201552119}%
  \BibitemOpen
  \bibfield  {author} {\bibinfo {author} {\bibfnamefont {H.}~\bibnamefont
  {Fehske}}, \bibinfo {author} {\bibfnamefont {G.}~\bibnamefont {Hager}}, \
  and\ \bibinfo {author} {\bibfnamefont {A.}~\bibnamefont {Pieper}},\ }\href
  {\doibase https://doi.org/10.1002/pssb.201552119} {\bibfield  {journal}
  {\bibinfo  {journal} {physica status solidi (b)}\ }\textbf {\bibinfo {volume}
  {252}},\ \bibinfo {pages} {1868} (\bibinfo {year} {2015})},\ \Eprint
  {http://arxiv.org/abs/https://onlinelibrary.wiley.com/doi/pdf/10.1002/pssb.201552119}
  {https://onlinelibrary.wiley.com/doi/pdf/10.1002/pssb.201552119} \BibitemShut
  {NoStop}%
\bibitem [{\citenamefont {Fonseca}\ \emph {et~al.}(2023)\citenamefont
  {Fonseca}, \citenamefont {Pereira},\ and\ \citenamefont
  {Barbosa}}]{PhysRevB.107.155432}%
  \BibitemOpen
  \bibfield  {author} {\bibinfo {author} {\bibfnamefont {D.~B.}\ \bibnamefont
  {Fonseca}}, \bibinfo {author} {\bibfnamefont {L.~F.~C.}\ \bibnamefont
  {Pereira}}, \ and\ \bibinfo {author} {\bibfnamefont {A.~L.}\ \bibnamefont
  {Barbosa}},\ }\href {\doibase 10.1103/PhysRevB.107.155432} {\bibfield
  {journal} {\bibinfo  {journal} {Physical Review B}\ }\textbf {\bibinfo
  {volume} {107}},\ \bibinfo {pages} {155432} (\bibinfo {year}
  {2023})}\BibitemShut {NoStop}%
\bibitem [{\citenamefont {Barthelemy}\ \emph {et~al.}(2008)\citenamefont
  {Barthelemy}, \citenamefont {Bertolotti},\ and\ \citenamefont
  {Wiersma}}]{bart}%
  \BibitemOpen
  \bibfield  {author} {\bibinfo {author} {\bibfnamefont {P.}~\bibnamefont
  {Barthelemy}}, \bibinfo {author} {\bibfnamefont {J.}~\bibnamefont
  {Bertolotti}}, \ and\ \bibinfo {author} {\bibfnamefont {D.~S.}\ \bibnamefont
  {Wiersma}},\ }\href {\doibase 10.1038/nature06948} {\bibfield  {journal}
  {\bibinfo  {journal} {Nature}\ }\textbf {\bibinfo {volume} {453}},\ \bibinfo
  {pages} {495} (\bibinfo {year} {2008})}\BibitemShut {NoStop}%
\bibitem [{\citenamefont {Groth}\ \emph {et~al.}(2012)\citenamefont {Groth},
  \citenamefont {Akhmerov},\ and\ \citenamefont
  {Beenakker}}]{PhysRevE.85.021138}%
  \BibitemOpen
  \bibfield  {author} {\bibinfo {author} {\bibfnamefont {C.~W.}\ \bibnamefont
  {Groth}}, \bibinfo {author} {\bibfnamefont {A.~R.}\ \bibnamefont {Akhmerov}},
  \ and\ \bibinfo {author} {\bibfnamefont {C.~W.~J.}\ \bibnamefont
  {Beenakker}},\ }\href {\doibase 10.1103/PhysRevE.85.021138} {\bibfield
  {journal} {\bibinfo  {journal} {Phys. Rev. E}\ }\textbf {\bibinfo {volume}
  {85}},\ \bibinfo {pages} {021138} (\bibinfo {year} {2012})}\BibitemShut
  {NoStop}%
\bibitem [{\citenamefont {Barthelemy}(2009)}]{barthelemy2009anomalous}%
  \BibitemOpen
  \bibfield  {author} {\bibinfo {author} {\bibfnamefont {P.}~\bibnamefont
  {Barthelemy}},\ }\emph {\bibinfo {title} {Anomalous Transport of Light}},\
  \href@noop {} {Ph.D. thesis},\ \bibinfo  {school} {Citeseer} (\bibinfo {year}
  {2009})\BibitemShut {NoStop}%
\bibitem [{\citenamefont {Brouwer}\ \emph {et~al.}(1997)\citenamefont
  {Brouwer}, \citenamefont {van Langen}, \citenamefont {Frahm}, \citenamefont
  {B\"uttiker},\ and\ \citenamefont {Beenakker}}]{PhysRevLett.79.913}%
  \BibitemOpen
  \bibfield  {author} {\bibinfo {author} {\bibfnamefont {P.~W.}\ \bibnamefont
  {Brouwer}}, \bibinfo {author} {\bibfnamefont {S.~A.}\ \bibnamefont {van
  Langen}}, \bibinfo {author} {\bibfnamefont {K.~M.}\ \bibnamefont {Frahm}},
  \bibinfo {author} {\bibfnamefont {M.}~\bibnamefont {B\"uttiker}}, \ and\
  \bibinfo {author} {\bibfnamefont {C.~W.~J.}\ \bibnamefont {Beenakker}},\
  }\href {\doibase 10.1103/PhysRevLett.79.913} {\bibfield  {journal} {\bibinfo
  {journal} {Phys. Rev. Lett.}\ }\textbf {\bibinfo {volume} {79}},\ \bibinfo
  {pages} {913} (\bibinfo {year} {1997})}\BibitemShut {NoStop}%
\bibitem [{\citenamefont {Amin}\ \emph {et~al.}(2018)\citenamefont {Amin},
  \citenamefont {Ray}, \citenamefont {Pal}, \citenamefont {Pandit},\ and\
  \citenamefont {Bid}}]{nature}%
  \BibitemOpen
  \bibfield  {author} {\bibinfo {author} {\bibfnamefont {K.~R.}\ \bibnamefont
  {Amin}}, \bibinfo {author} {\bibfnamefont {S.~S.}\ \bibnamefont {Ray}},
  \bibinfo {author} {\bibfnamefont {N.}~\bibnamefont {Pal}}, \bibinfo {author}
  {\bibfnamefont {R.}~\bibnamefont {Pandit}}, \ and\ \bibinfo {author}
  {\bibfnamefont {A.}~\bibnamefont {Bid}},\ }\href {\doibase
  10.1038/s42005-017-0001-4} {\bibfield  {journal} {\bibinfo  {journal}
  {Communications Physics}\ }\textbf {\bibinfo {volume} {1}} (\bibinfo {year}
  {2018}),\ 10.1038/s42005-017-0001-4}\BibitemShut {NoStop}%
\bibitem [{\citenamefont {Pessoa}\ \emph {et~al.}(2021)\citenamefont {Pessoa},
  \citenamefont {Barbosa}, \citenamefont {Vasconcelos},\ and\ \citenamefont
  {Macedo}}]{PhysRevE.104.054129}%
  \BibitemOpen
  \bibfield  {author} {\bibinfo {author} {\bibfnamefont {N.~L.}\ \bibnamefont
  {Pessoa}}, \bibinfo {author} {\bibfnamefont {A.~L.~R.}\ \bibnamefont
  {Barbosa}}, \bibinfo {author} {\bibfnamefont {G.~L.}\ \bibnamefont
  {Vasconcelos}}, \ and\ \bibinfo {author} {\bibfnamefont {A.~M.~S.}\
  \bibnamefont {Macedo}},\ }\href {\doibase 10.1103/PhysRevE.104.054129}
  {\bibfield  {journal} {\bibinfo  {journal} {Phys. Rev. E}\ }\textbf {\bibinfo
  {volume} {104}},\ \bibinfo {pages} {054129} (\bibinfo {year}
  {2021})}\BibitemShut {NoStop}%
\bibitem [{\citenamefont {Barbosa}\ \emph {et~al.}(2022)\citenamefont
  {Barbosa}, \citenamefont {de~Lima}, \citenamefont {Gonz\'alez}, \citenamefont
  {Pessoa}, \citenamefont {Mac\^edo},\ and\ \citenamefont
  {Vasconcelos}}]{PhysRevLett.128.236803}%
  \BibitemOpen
  \bibfield  {author} {\bibinfo {author} {\bibfnamefont {A.~L.~R.}\
  \bibnamefont {Barbosa}}, \bibinfo {author} {\bibfnamefont {T.~H.~V.}\
  \bibnamefont {de~Lima}}, \bibinfo {author} {\bibfnamefont {I.~R.~R.}\
  \bibnamefont {Gonz\'alez}}, \bibinfo {author} {\bibfnamefont {N.~L.}\
  \bibnamefont {Pessoa}}, \bibinfo {author} {\bibfnamefont {A.~M.~S.}\
  \bibnamefont {Mac\^edo}}, \ and\ \bibinfo {author} {\bibfnamefont {G.~L.}\
  \bibnamefont {Vasconcelos}},\ }\href {\doibase
  10.1103/PhysRevLett.128.236803} {\bibfield  {journal} {\bibinfo  {journal}
  {Phys. Rev. Lett.}\ }\textbf {\bibinfo {volume} {128}},\ \bibinfo {pages}
  {236803} (\bibinfo {year} {2022})}\BibitemShut {NoStop}%
\bibitem [{\citenamefont {Zhao}\ \emph {et~al.}(2017)\citenamefont {Zhao},
  \citenamefont {Li}, \citenamefont {Yang}, \citenamefont {Han}, \citenamefont
  {Su},\ and\ \citenamefont {Zou}}]{Zhao_2017}%
  \BibitemOpen
  \bibfield  {author} {\bibinfo {author} {\bibfnamefont {L.}~\bibnamefont
  {Zhao}}, \bibinfo {author} {\bibfnamefont {W.}~\bibnamefont {Li}}, \bibinfo
  {author} {\bibfnamefont {C.}~\bibnamefont {Yang}}, \bibinfo {author}
  {\bibfnamefont {J.}~\bibnamefont {Han}}, \bibinfo {author} {\bibfnamefont
  {Z.}~\bibnamefont {Su}}, \ and\ \bibinfo {author} {\bibfnamefont
  {Y.}~\bibnamefont {Zou}},\ }\href {\doibase 10.1371/journal.pone.0170467}
  {\bibfield  {journal} {\bibinfo  {journal} {PLOS ONE}\ }\textbf {\bibinfo
  {volume} {12}},\ \bibinfo {pages} {e0170467} (\bibinfo {year}
  {2017})}\BibitemShut {NoStop}%
\bibitem [{\citenamefont {Bao}\ \emph {et~al.}(2021)\citenamefont {Bao},
  \citenamefont {Zhang}, \citenamefont {Zhang}, \citenamefont {Wu},
  \citenamefont {Luo}, \citenamefont {Zhou}, \citenamefont {Li}, \citenamefont
  {Hou}, \citenamefont {Yao}, \citenamefont {Liu}, \citenamefont {Yu},
  \citenamefont {Li}, \citenamefont {Duan}, \citenamefont {Yao}, \citenamefont
  {Wang},\ and\ \citenamefont {Zhou}}]{PhysRevLett.126.206804}%
  \BibitemOpen
  \bibfield  {author} {\bibinfo {author} {\bibfnamefont {C.}~\bibnamefont
  {Bao}}, \bibinfo {author} {\bibfnamefont {H.}~\bibnamefont {Zhang}}, \bibinfo
  {author} {\bibfnamefont {T.}~\bibnamefont {Zhang}}, \bibinfo {author}
  {\bibfnamefont {X.}~\bibnamefont {Wu}}, \bibinfo {author} {\bibfnamefont
  {L.}~\bibnamefont {Luo}}, \bibinfo {author} {\bibfnamefont {S.}~\bibnamefont
  {Zhou}}, \bibinfo {author} {\bibfnamefont {Q.}~\bibnamefont {Li}}, \bibinfo
  {author} {\bibfnamefont {Y.}~\bibnamefont {Hou}}, \bibinfo {author}
  {\bibfnamefont {W.}~\bibnamefont {Yao}}, \bibinfo {author} {\bibfnamefont
  {L.}~\bibnamefont {Liu}}, \bibinfo {author} {\bibfnamefont {P.}~\bibnamefont
  {Yu}}, \bibinfo {author} {\bibfnamefont {J.}~\bibnamefont {Li}}, \bibinfo
  {author} {\bibfnamefont {W.}~\bibnamefont {Duan}}, \bibinfo {author}
  {\bibfnamefont {H.}~\bibnamefont {Yao}}, \bibinfo {author} {\bibfnamefont
  {Y.}~\bibnamefont {Wang}}, \ and\ \bibinfo {author} {\bibfnamefont
  {S.}~\bibnamefont {Zhou}},\ }\href {\doibase 10.1103/PhysRevLett.126.206804}
  {\bibfield  {journal} {\bibinfo  {journal} {Phys. Rev. Lett.}\ }\textbf
  {\bibinfo {volume} {126}},\ \bibinfo {pages} {206804} (\bibinfo {year}
  {2021})}\BibitemShut {NoStop}%
\bibitem [{\citenamefont {Ulybyshev}\ \emph {et~al.}(2013)\citenamefont
  {Ulybyshev}, \citenamefont {Buividovich}, \citenamefont {Katsnelson},\ and\
  \citenamefont {Polikarpov}}]{PhysRevLett.111.056801}%
  \BibitemOpen
  \bibfield  {author} {\bibinfo {author} {\bibfnamefont {M.~V.}\ \bibnamefont
  {Ulybyshev}}, \bibinfo {author} {\bibfnamefont {P.~V.}\ \bibnamefont
  {Buividovich}}, \bibinfo {author} {\bibfnamefont {M.~I.}\ \bibnamefont
  {Katsnelson}}, \ and\ \bibinfo {author} {\bibfnamefont {M.~I.}\ \bibnamefont
  {Polikarpov}},\ }\href {\doibase 10.1103/PhysRevLett.111.056801} {\bibfield
  {journal} {\bibinfo  {journal} {Phys. Rev. Lett.}\ }\textbf {\bibinfo
  {volume} {111}},\ \bibinfo {pages} {056801} (\bibinfo {year}
  {2013})}\BibitemShut {NoStop}%
\bibitem [{\citenamefont {Kantelhardt}\ \emph {et~al.}(2002)\citenamefont
  {Kantelhardt}, \citenamefont {Zschiegner}, \citenamefont {Koscielny-Bunde},
  \citenamefont {Havlin}, \citenamefont {Bunde},\ and\ \citenamefont
  {Stanley}}]{kantelhardt2002multifractal}%
  \BibitemOpen
  \bibfield  {author} {\bibinfo {author} {\bibfnamefont {J.~W.}\ \bibnamefont
  {Kantelhardt}}, \bibinfo {author} {\bibfnamefont {S.}~\bibnamefont
  {Zschiegner}}, \bibinfo {author} {\bibfnamefont {E.}~\bibnamefont
  {Koscielny-Bunde}}, \bibinfo {author} {\bibfnamefont {S.}~\bibnamefont
  {Havlin}}, \bibinfo {author} {\bibfnamefont {A.}~\bibnamefont {Bunde}}, \
  and\ \bibinfo {author} {\bibfnamefont {H.~E.}\ \bibnamefont {Stanley}},\
  }\href {\doibase 10.1016/s0378-4371(02)01383-3} {\bibfield  {journal}
  {\bibinfo  {journal} {Physica A: Statistical Mechanics and Its Applications}\
  }\textbf {\bibinfo {volume} {316}},\ \bibinfo {pages} {87} (\bibinfo {year}
  {2002})}\BibitemShut {NoStop}%
\bibitem [{\citenamefont {Datta}(2005)}]{datta}%
  \BibitemOpen
  \bibfield  {author} {\bibinfo {author} {\bibfnamefont {S.}~\bibnamefont
  {Datta}},\ }\href@noop {} {\emph {\bibinfo {title} {Quantum transport: atom
  to transistor}}}\ (\bibinfo  {publisher} {Cambridge university press},\
  \bibinfo {year} {2005})\BibitemShut {NoStop}%
\bibitem [{\citenamefont {Groth}\ \emph {et~al.}(2014)\citenamefont {Groth},
  \citenamefont {Wimmer}, \citenamefont {Akhmerov},\ and\ \citenamefont
  {Waintal}}]{kwant}%
  \BibitemOpen
  \bibfield  {author} {\bibinfo {author} {\bibfnamefont {C.~W.}\ \bibnamefont
  {Groth}}, \bibinfo {author} {\bibfnamefont {M.}~\bibnamefont {Wimmer}},
  \bibinfo {author} {\bibfnamefont {A.~R.}\ \bibnamefont {Akhmerov}}, \ and\
  \bibinfo {author} {\bibfnamefont {X.}~\bibnamefont {Waintal}},\ }\href@noop
  {} {\bibfield  {journal} {\bibinfo  {journal} {New Journal of Physics}\
  }\textbf {\bibinfo {volume} {16}},\ \bibinfo {pages} {063065} (\bibinfo
  {year} {2014})}\BibitemShut {NoStop}%
\bibitem [{\citenamefont {Cysne}\ \emph
  {et~al.}(2018{\natexlab{a}})\citenamefont {Cysne}, \citenamefont {Ferreira},\
  and\ \citenamefont {Rappoport}}]{PhysRevB.98.045407}%
  \BibitemOpen
  \bibfield  {author} {\bibinfo {author} {\bibfnamefont {T.~P.}\ \bibnamefont
  {Cysne}}, \bibinfo {author} {\bibfnamefont {A.}~\bibnamefont {Ferreira}}, \
  and\ \bibinfo {author} {\bibfnamefont {T.~G.}\ \bibnamefont {Rappoport}},\
  }\href {\doibase 10.1103/PhysRevB.98.045407} {\bibfield  {journal} {\bibinfo
  {journal} {Phys. Rev. B}\ }\textbf {\bibinfo {volume} {98}},\ \bibinfo
  {pages} {045407} (\bibinfo {year} {2018}{\natexlab{a}})}\BibitemShut
  {NoStop}%
\bibitem [{\citenamefont {Barbosa}\ \emph {et~al.}(2021)\citenamefont
  {Barbosa}, \citenamefont {Ramos},\ and\ \citenamefont
  {Ferreira}}]{PhysRevB.103.L081111}%
  \BibitemOpen
  \bibfield  {author} {\bibinfo {author} {\bibfnamefont {A.~L.~R.}\
  \bibnamefont {Barbosa}}, \bibinfo {author} {\bibfnamefont {J.~G. G.~S.}\
  \bibnamefont {Ramos}}, \ and\ \bibinfo {author} {\bibfnamefont
  {A.}~\bibnamefont {Ferreira}},\ }\href {\doibase
  10.1103/PhysRevB.103.L081111} {\bibfield  {journal} {\bibinfo  {journal}
  {Phys. Rev. B}\ }\textbf {\bibinfo {volume} {103}},\ \bibinfo {pages}
  {L081111} (\bibinfo {year} {2021})}\BibitemShut {NoStop}%
\bibitem [{\citenamefont {Cysne}\ \emph
  {et~al.}(2018{\natexlab{b}})\citenamefont {Cysne}, \citenamefont {Garcia},
  \citenamefont {Rocha},\ and\ \citenamefont {Rappoport}}]{PhysRevB.97.085413}%
  \BibitemOpen
  \bibfield  {author} {\bibinfo {author} {\bibfnamefont {T.~P.}\ \bibnamefont
  {Cysne}}, \bibinfo {author} {\bibfnamefont {J.~H.}\ \bibnamefont {Garcia}},
  \bibinfo {author} {\bibfnamefont {A.~R.}\ \bibnamefont {Rocha}}, \ and\
  \bibinfo {author} {\bibfnamefont {T.~G.}\ \bibnamefont {Rappoport}},\ }\href
  {\doibase 10.1103/PhysRevB.97.085413} {\bibfield  {journal} {\bibinfo
  {journal} {Phys. Rev. B}\ }\textbf {\bibinfo {volume} {97}},\ \bibinfo
  {pages} {085413} (\bibinfo {year} {2018}{\natexlab{b}})}\BibitemShut
  {NoStop}%
\bibitem [{\citenamefont {Castro~Neto}\ \emph {et~al.}(2009)\citenamefont
  {Castro~Neto}, \citenamefont {Guinea}, \citenamefont {Peres}, \citenamefont
  {Novoselov},\ and\ \citenamefont {Geim}}]{RevModPhys.81.109}%
  \BibitemOpen
  \bibfield  {author} {\bibinfo {author} {\bibfnamefont {A.~H.}\ \bibnamefont
  {Castro~Neto}}, \bibinfo {author} {\bibfnamefont {F.}~\bibnamefont {Guinea}},
  \bibinfo {author} {\bibfnamefont {N.~M.~R.}\ \bibnamefont {Peres}}, \bibinfo
  {author} {\bibfnamefont {K.~S.}\ \bibnamefont {Novoselov}}, \ and\ \bibinfo
  {author} {\bibfnamefont {A.~K.}\ \bibnamefont {Geim}},\ }\href {\doibase
  10.1103/RevModPhys.81.109} {\bibfield  {journal} {\bibinfo  {journal} {Rev.
  Mod. Phys.}\ }\textbf {\bibinfo {volume} {81}},\ \bibinfo {pages} {109}
  (\bibinfo {year} {2009})}\BibitemShut {NoStop}%
\bibitem [{\citenamefont {Zaburdaev}\ \emph {et~al.}(2015)\citenamefont
  {Zaburdaev}, \citenamefont {Denisov},\ and\ \citenamefont
  {Klafter}}]{zaburdaev2015levy}%
  \BibitemOpen
  \bibfield  {author} {\bibinfo {author} {\bibfnamefont {V.}~\bibnamefont
  {Zaburdaev}}, \bibinfo {author} {\bibfnamefont {S.}~\bibnamefont {Denisov}},
  \ and\ \bibinfo {author} {\bibfnamefont {J.}~\bibnamefont {Klafter}},\
  }\href@noop {} {\bibfield  {journal} {\bibinfo  {journal} {Reviews of Modern
  Physics}\ }\textbf {\bibinfo {volume} {87}},\ \bibinfo {pages} {483}
  (\bibinfo {year} {2015})}\BibitemShut {NoStop}%
\bibitem [{\citenamefont {Zhai}\ and\ \citenamefont
  {Xu}(2005)}]{PhysRevLett.94.246601}%
  \BibitemOpen
  \bibfield  {author} {\bibinfo {author} {\bibfnamefont {F.}~\bibnamefont
  {Zhai}}\ and\ \bibinfo {author} {\bibfnamefont {H.~Q.}\ \bibnamefont {Xu}},\
  }\href {\doibase 10.1103/PhysRevLett.94.246601} {\bibfield  {journal}
  {\bibinfo  {journal} {Phys. Rev. Lett.}\ }\textbf {\bibinfo {volume} {94}},\
  \bibinfo {pages} {246601} (\bibinfo {year} {2005})}\BibitemShut {NoStop}%
\bibitem [{\citenamefont {Gnutzmann}\ and\ \citenamefont
  {Seif}(2004)}]{PhysRevE.69.056219}%
  \BibitemOpen
  \bibfield  {author} {\bibinfo {author} {\bibfnamefont {S.}~\bibnamefont
  {Gnutzmann}}\ and\ \bibinfo {author} {\bibfnamefont {B.}~\bibnamefont
  {Seif}},\ }\href {\doibase 10.1103/PhysRevE.69.056219} {\bibfield  {journal}
  {\bibinfo  {journal} {Phys. Rev. E}\ }\textbf {\bibinfo {volume} {69}},\
  \bibinfo {pages} {056219} (\bibinfo {year} {2004})}\BibitemShut {NoStop}%
\end{thebibliography}%

\end{document}